\documentclass{elsarticle}

\usepackage{hyperref}

\journal{{\tt arXiv}, produced by using the {\tt elsarticle} class.}
\usepackage{amsmath,amssymb}
\usepackage{bm}
\usepackage{graphicx}
\usepackage{color}

\usepackage{setspace}

\newtheorem{theorem}{Theorem}
\newtheorem{proposition}[theorem]{Proposition}
\newtheorem{definition}[theorem]{Definition}
\newtheorem{lemma}[theorem]{Lemma}

\let\epsilon\varepsilon
\newcommand{\eps}{\varepsilon}

\definecolor{myred}{rgb}{1,0,0}
\definecolor{mygray}{rgb}{0.5,0.5,0.5}
\newcommand\gra[1]{\textcolor{mygray}{#1}}
\newcommand\red[1]{\textcolor{myred}{#1}}

\newcommand{\atopp}[2]{\genfrac{}{}{0pt}{}{#1}{#2}}

\newcommand\vr{{\bm r}}

\def\smfrac#1#2{{\textstyle\frac{#1}{#2}}}

\newcommand{\be}{\begin{equation}}
\newcommand{\ee}{\end{equation}}
\newcommand{\bes}{\begin{equation*}}
\newcommand{\ees}{\end{equation*}}
\newcommand{\ef}[1]{\, #1}

\DeclareMathOperator*{\argmax}{argmax}

\newcommand{\FG}{\mathrm{FG}}
\newcommand{\mmin}{m_{\rm min}}

\newcommand{\ka}{\kappa}

\newcommand{\bxi}{\xi}

\newcommand{\cX}{\mathcal{X}}
\newcommand{\cY}{\mathcal{Y}}

\newcommand{\cA}{\mathcal{A}}

\newcommand{\cC}{\mathcal{C}}

\newcommand{\eee}{\mathbb{E}}

\newcommand{\lam}{\lambda}

\newcommand{\myrulino}{\raisebox{-3pt}{\rule{.5pt}{5.5pt}}}

\begin{document}

\begin{frontmatter}

\title{The complexity of the Multiple Pattern Matching
%\rule{0pt}{14pt}%
Problem for random strings}
% \tnoteref{mytitlenote}}
% \tnotetext[mytitlenote]{Fully documented templates are available in the elsarticle package on \href{http://www.ctan.org/tex-archive/macros/latex/contrib/elsarticle}{CTAN}.}

\author{Fr\'ed\'erique Bassino\fnref{myfootnote1}}
\author{Tsinjo Rakotoarimalala\fnref{myfootnote2}}
\author{Andrea Sportiello\fnref{myfootnote3}}

\address{Universit\'e Paris 13, Sorbonne Paris Cit\'e, LIPN, CNRS UMR 7030,\\
99 av.\ J.-B.\ Cl\'ement, F-93430 Villetaneuse, France}
\fntext[myfootnote1]{\tt bassin\makebox[0pt][l]{o}o@l\makebox[0pt][l]{i}ipn.un\makebox[0pt][l]{i}iv-paris13.fr}
\fntext[myfootnote2]{\tt rakotoarimalal\makebox[0pt][l]{a}a@l\makebox[0pt][l]{i}ipn.un\makebox[0pt][l]{i}iv-paris13.fr}
\fntext[myfootnote3]{\tt sport\makebox[0pt][l]{i}iello@l\makebox[0pt][l]{i}ipn.un\makebox[0pt][l]{i}iv-paris13.fr}


%% \author[mysecondaryaddress]{Global Customer Service\corref{mycorrespondingauthor}}
%% \cortext[mycorrespondingauthor]{Corresponding author}
%% \ead{support@elsevier.com}

%% \address[mymainaddress]{1600 John F Kennedy Boulevard, Philadelphia}
%% \address[mysecondaryaddress]{360 Park Avenue South, New York}

\begin{abstract}
We generalise a multiple string pattern
  matching algorithm, recently proposed by Fredriksson and Grabowski
  [J.\ Discr.\ Alg.\ 7, 2009], to deal with arbitrary dictionaries on
  an alphabet of size $s$. If $r_m$ is the number of words of length
  $m$ in the dictionary, and
  $\phi(r) = \max_m \ln(s\, m\, r_m)/m$, the complexity rate for the
  string characters to be read by this algorithm is at most
  $\kappa_{{}_\textrm{\tiny UB}}\, \phi(r)$ for some constant
  $\kappa_{{}_\textrm{\tiny UB}}$.\\ \rule{20pt}{0pt}On the other
  side, we generalise the classical lower bound of Yao [SIAM
    J.\ Comput.\ 8, 1979], for the problem with a single pattern, to
  deal with arbitrary dictionaries, and determine it to be at least
  $\kappa_{{}_\textrm{\tiny LB}}\, \phi(r)$. This proves the
  optimality of the algorithm, improving and correcting previous
  claims.
\end{abstract}


\end{frontmatter}

% \linenumbers

% VERSION OF ABSTRACT FIRST SUBMITTED TO SODA WEB INTERFACE
%% A pattern-matching algorithm is an algorithm that, for a given set of
%% words (dictionary), and a string of length $n$, identifies all
%% occurrences of the words in the string. Its average complexity, for a
%% given dictionary, is the limit for $n \to \infty$ of the fraction of
%% symbols in the string that need to be read, averaged over uniform
%% random strings.
%% This paper presents a generalisation of an algorithm of Fredriksson
%% and Grabowski, to deal with arbitrary dictionaries (beyond the case of
%% words of a unique length). We analyse the complexity of the given
%% algorithm, and, following ideas of Yao, provide a lower bound for the
%% problem. We can prove that both the algorithmic upper bound and the
%% theoretical lower bound are within a constant factor to a simple
%% parameter $\phi$ associated to the dictionary, namely, if there are
%% $r_m$ words of length $m$, $\phi=\max_m \ln(r_m m)/m$. This proves the
%% optimality of the algorithm.}

% \vspace{2mm}

%%%%%%%%%%%%%%%%%%%%%%%%%%%%%%%%%%%%%%%%%%%%%%%%%%%%%%%
% \mysection{1. The problem}
\section{The problem}
% \label{sec.intro}
%%%%%%%%%%%%%%%%%%%%%%%%%%%%%%%%%%%%%%%%%%%%%%%%%%%%%%%

\subsection{Definition of the problem}

Let $\Sigma$ be an alphabet of $s$ symbols, 
$\bxi=(\xi_1,\ldots,\xi_n) \in \Sigma^n$ a word of $n$ characters (the
input \emph{text string}), $D=\{w_1,\ldots,w_k\}$, $w_i \in \Sigma^*$
a collection of words (the \emph{dictionary}). We say that $w$ occurs
in $\bxi$ at position $j$ if 
$w \equiv \xi_j \xi_{j+1} \cdots \xi_{j+|w|-1}$.  Let $r_m(D)$ be the
number of words of length $m$ in $D$.  We call 
$\vr(D)=\{r_m(D)\}_{m \geq 1}$ the \emph{content} of $D$, a notion of
crucial importance in this paper. 

The \emph{multiple string pattern matching problem} (MPMP), for the
datum $(D,\bxi)$, is the problem of identifying all the occurrences of
the words in $D$ inside the text $\bxi$ (cf.\ Figure~\ref{fig.exa}).

A version of the problem can be defined in the case of infinite
dictionaries ($k=\infty$), where, as we may assume that all the words
of the dictionary are distinct, we can suppose that 
$r_m \leq s^m$. The analysis of the present paper treats in an unified
way the case of finite and infinite dictionaries.

%% This is a classical and well-studied problem, mathematically
%% interesting and, obviously, with several applications in Computer
%% Science (for example, .........).  

The first seminal works have concerned single-word dictionaries.
Results included the determination of the average complexity of the
problem, the design of efficient algorithms (notably
Knuth-Morris-Pratt and Boyer-Moore), and have led to the far-reaching
definition of Aho-Corasick automata \cite{ahoco,BM,KMP,yao}.

% CITE HERE ALSO CHENG MARR ????? TALK AT ALL ABOUT APPROX MATCHING ????

\begin{figure}[tb]
\begin{center}
{\small
\begin{tabular}{r|l|l}
$\bxi$
              & 
%\makebox[0pt][c]{\myrulino}\makebox[0pt][c]{${}^0$}\phantom{\tt 11001}%
\rule{23.5pt}{0pt}%
\makebox[0pt][c]{\myrulino}\makebox[0pt][c]{${}^5$}\phantom{\tt 01011}%
\makebox[0pt][c]{\myrulino}\makebox[0pt][c]{${}^{10}$}\phantom{\tt 00100}%
\makebox[0pt][c]{\myrulino}\makebox[0pt][c]{${}^{15}$}\phantom{\tt 00100}%
\makebox[0pt][c]{\myrulino}\makebox[0pt][c]{${}^{20}$}\phantom{\tt 10101}%
\makebox[0pt][c]{\myrulino}\makebox[0pt][c]{${}^{25}$}\phantom{\tt 10101}%
\makebox[0pt][c]{\myrulino}\makebox[0pt][c]{${}^{30}$}\phantom{\tt 01000}%
\makebox[0pt][c]{\myrulino}\makebox[0pt][c]{${}^{35}$}\phantom{\tt 11100}%
\makebox[0pt][c]{\myrulino}\makebox[0pt][c]{${}^{40}$}\phantom{\tt 10110}%
\makebox[0pt][c]{\myrulino}\makebox[0pt][c]{${}^{45}$}\\
%\phantom{\tt 01110}%
%\makebox[0pt][c]{\myrulino}\makebox[0pt][c]{${}^{50}$}\\
$D$ \rule{20pt}{0pt}
              & {\tt 110010101100100001001010110101010001110010110}
& \multicolumn{1}{|c}{output}
\\
%01110} \\
\hline
   {\tt 0010} & {\tt ..*---....*---.*--*---................*---...} &
%.....} &
$3, 11, 16, 19, 39$\\
  {\tt 01010} & {\tt ...*----...........*----..*-*----............} &
% .....} &
$4, 20, 27, 29$\\
\makebox[0pt][r]{\tt 1011001} & {\tt ......*------................................} &
% *------...} &
7 \\
%, 41 \\
\hline
%%    {\tt 0010} & \rule{0pt}{12pt}3, 11, 16, 19, 39\\
%%   {\tt 01010} & 4, 20, 27, 29\\
%% {\tt 1011001} & 7, 41 \\
%% \hline
\end{tabular}
}
\end{center}
\caption{Typical output of a multiple string pattern matching
  problem. In this case $s=2$
% $D=\{{\tt 0010}, {\tt 01010}, {\tt 1011001} \}$, 
and $\vr(D)=(0,0,0,1,1,0,1,0,\ldots)$.
\label{fig.exa}}
%% , for the dictionary 
%% $D=\{
%%    {\tt 0010},
%%   {\tt 01010},
%% {\tt 1011001} \}$, 
%% %
%% and a text of size $n=50$. The occurrences of a word $w$ have been
%% marked with a {\tt *} symbol. The desired outcome is the collection of
%% lists shown in the bottom part.  Note how different words and
%% different occurrences of the same word may overlap.\label{fig.exa}}
\end{figure}

The computational complexity of the problem, for a given dictionary
$D$, in worst case among texts $\xi$ of length $n$, is a problem
solved by Rivest long ago~\cite{rivest}, with a deceivingly simple
answer (within our complexity paradigm, based on text accesses, see
below): for each word, there exist texts that need to be read
entirely.
% (within our complexity paradigm, based on text accesses, see below).  

On the other side, the \emph{average-case} analysis, over
texts $\xi$ sampled uniformly in $\Sigma^n$ (in the limit of large
$n$) is an interesting problem, and its determination for any given
dictionary $D$ is the ideal goal of this paper.

% been solved by Rivest long ago~\cite{rivest}.

Unfortunately, the \emph{exact} determination of the complexity for an
arbitrary dictionary seems a formidable task, and we have to revert to
a more concrete challenge. More precisely, our goal is as follows.
\begin{itemize}
\item 
Analogously to what is done in Knuth, Morris and Pratt \cite{KMP} and
in Yao \cite{yao}, we do not analyse the time complexity, but rather
the complexity in terms of \emph{character accesses} to original text.
\item
The asymptotic complexity is $\Theta(n)$, in other words there exists
some constant $\Phi(D)$ such that the quantity of interest is
$\Phi(D)\, n + o(n)$.  Again, analogously to previous literature,
instead of determining the quantity $\Phi(D)$ exactly, we determine
the \emph{functional dependence} of $\Phi$ from $D$, up to a finite
multiplicative constant, and thus produce upper and lower bounds whose
ratio is uniformly bounded.
\end{itemize}
We now describe in more detail the subtleties of the forementioned
complexity paradigm.

Let $\mmin$ be the length of the shortest pattern in $D$. In
particular, the number of character accesses is at most $n$, the case
in which the whole text is read, and is $n-\mathcal{O}(1)$ in worst
case.\footnote{Think to the case in which $\xi$ is composed of a
  concatenation of words from $D$, possibly up to the last $k$
  characters, with $k < \mmin$.} It is also at least
$n/\mmin+\mathcal{O}(1)$, because one needs to read at least one
character out of $\mmin$ consecutive ones. Another trivial consequence
is that we can restrict to the case \hbox{$\mmin \geq 2$}, as, if we have
even a single word in $D$ of length 1, we know in advance that the
whole text needs to be read.

The average number of characters to be read for texts of length $n$ is
a super-additive sequence (because solving an instance $\xi$ of size
$n_1 + n_2$ is harder than solving the two instances, of sizes $n_1$
and $n_2$, associated to the appropriate prefix and suffix of
$\xi$). As a consequence of Fekete's Subadditive Lemma, the limit
fraction of required character comparisons exists and is equal to the
lim-sup of the same quantity. The Fekete's Lemma in itself leaves open
the possibility that the limit is infinite, however from the
forementioned trivial upper bound we can conclude that this limit is a
finite constant associated to the dictionary. 

We introduce the quantity $\Phi(D)$, by defining this fraction to be
$\Phi(D)/\ln s$, and we have thus determined that
\[
\frac{\ln s}{\mmin(D)}
 \leq \Phi(D) \leq 
\ln s
\ef.
\]
The presence of the logarithmic prefactor is a useful convention,
associated to the fact that reading $N$ characters in a $s$-symbol
alphabet gives a $N \log_2(s)$ amount of Shannon information entropy
(we further use natural basis for logarithms, in order to simplify the
calculations). An advantage of the rescaled quantity is the fact that
it has a form of stability under reduction of the text into $q$-grams
(i.e., a text of length $qn$ on an alphabet $\Sigma$ of $s$ symbols
has the same Shannon entropy, $(n q) \log_2(s) = n \log_2(s^q)$, of
the text of length $n$ consituted of symbols in $\Sigma^q$).  Our
ideal task would be the determination of $\Phi(D)$.

% \item 
Yao \cite{yao} determines in particular that, if $D$ consists of a
single word of length $m$, when $1 \ll m \ll n$, the complexity above
is given by a certain expression in $m$ and $s$, up to a finite
multiplicative constant, for \emph{almost all patterns}, i.e.\ for
almost all of the $s^m$ possible words of the given length. Patterns
with a \emph{different} complexity, in fact, always have a
\emph{smaller} complexity, i.e.\ it is not excluded that there are a
few atypically-simpler words, while it is established that there are
no (significantly) atypically-harder ones. In analogy to this result,
it is natural to imagine that, as we will see in the following, almost
all dictionaries with the same content $\vr$ have the same complexity
up to finite multiplicative factors, and the few remaining ones have a
smaller complexity. For this reason we find interesting to define
\begin{align}
\Phi_{\rm min}(\vr)&:=\min_{D} \Phi(D) 
\ef;
\\
\Phi_{\rm aver}(\vr)&:=\eee_D \big( \Phi(D) \big)
\ef;
\\
\Phi_{\rm max}(\vr)&:=\max_{D} \Phi(D) 
\ef;
\end{align}
where min, max and average are taken over $D$ such that $\vr(D)=\vr$,
and one of our aims is to prove that $\Phi_{\rm aver}(\vr)$ and
$\Phi_{\rm max}(\vr)$ functions have the same behaviour up to a
multiplicative constant (while this is not the case for 
$\Phi_{\rm min}(\vr)$).

% \item 
Also the exact determination of $\Phi_{\rm aver}(\vr)$ and $\Phi_{\rm
  max}(\vr)$ is an overwhelming task. We will content ourself of a
determination of these functions up to a multiplicative constant,
i.e.\ the identification of a function $\phi(\vr)$, and a constant
$\kappa$, such that
\be
\frac{\phi(\vr)}{\kappa}
\leq 
\Phi_{\rm aver}(\vr) \leq
\Phi_{\rm max}(\vr)
% \Phi(\vr) 
\leq 
\kappa \, \phi(\vr)
\ef.
\label{eq.boukk}
\ee
Yet again, this is not dissimilar to what is done in the seminal Yao
paper. 

%% The `true' averages are done by assuming that there are no repeated
%% words. Through this assumptions, we can derive in particular that 
%% %
%% $r_m \leq s^m$ without loss of generality. But then, when we state
%% that $\Phi_{\rm aver}(\vr) \leq \Phi'_{\rm aver}(\vr)$ and we perform
%% the analysis on the case in which the repetitions are allowed, we keep
%% on assuming that $r_m \leq s^m$ (even if this is not implied in the
%% model with repetitions), because we compare systems with the same
%% $\{r_m\}$, and our ultimate interest is in the model without
%% repetitions.

Note that the average in $\Phi_{\rm aver}(\vr)$ allows for the
repetition of the same word in the dictionary, and for the presence of
pairs of words in the dictionary which are one factor of the
other. These problems are thus trivially reduced to problems on a
smaller dictionary.  If we call $\Phi'_{\rm aver}(\vr)$ the average
performed while excluding these degenerate cases, from the
monotonicity of the complexity as a function of $D$
(w.r.t.\ inclusion) we get that 
$\Phi_{\rm aver}(\vr) \leq \Phi'_{\rm aver}(\vr) \leq \Phi_{\rm
  max}(\vr)$, so that our theorem, as a corollary, establishes the
behaviour of the more interesting quantity $\Phi'_{\rm aver}(\vr)$.
Observe that it is legitimate to do this even while still keeping the
bound $r_m \leq s^m$, which occurs only in problems without
repetitions, because we are comparing the two types of averages while
the content $\vr$ is kept fixed.

We shall comment on the fact that finding the complexity up to a
multiplicative constant is not an easy task \emph{a priori}. Suppose
that we have a concrete working strategy to determine an upper bound
for a given vector $\vr$, and a second strategy to determine a lower
bound for given $\vr$. Then one may na\"ively think that $\kappa$ is
related to the maximum, over all $\vr$, of the associated ratio. This
idea is wrong, because the space of possible $\vr$ is not a compact,
and the `maximum' of this ratio is in fact a `supremum', which may
well be infinity. So, any valid lower and upper bounds provide an
estimate of the complexity for a given content $\vr$ up to a constant,
but we need ``good'' bound strategies in order to capture the full
functional dependence of the complexity from the content parameters
$\vr=\{r_m\}$.

\subsection{The result and its implications}
%-------------------------------------------------------

Let
\be
\label{eq.phir}
\phi(\vr):=\max_m \frac{1}{m} \ln (s\, m \, r_m)
\ef.
\ee
Note that, even in the case of infinite dictionaries $D$, with all
words of length at least $\mmin$, $\phi(\vr(D))$ must evaluate to a
finite value, as it is bounded by 
$\ln s + \frac{1}{\mmin} \ln(s\,\mmin)$
(because $r_m \leq s^m$).

Our aim is to prove the following theorem, valid in the interesting
regime $s \geq 2$ and $\mmin \geq 2$.
\begin{theorem}
\label{thm1}
Let $\ka_s = 5 \frac{\sqrt{s}}{\ln s}$.
For all contents $\vr$, the complexity of the MPMP on an
alphabet of size $s$ satisfies the bounds
\be
\frac{1}{\ka_s} 
\left( \phi(\vr) + \frac{1}{2s\, \mmin} \right)
\leq \Phi_{\rm aver}(\vr) \leq \Phi_{\rm max}(\vr)
\leq 2 
\left( \phi(\vr) + \frac{1}{2s\, \mmin} \right)
\ef.
\label{eq.3654635}
\ee
\end{theorem}
%% The values of $\ka_s$ are determined as solution of a transcendental
%% equation discussed in Section \ref{sec.lb}. The first few values are
%% \[
%% \begin{array}{c|c}
%% s & \ka_s \\
%% \hline
%%  2 &   7.389184 \\
%%  3 &   8.382134 \\
%%  4 &  10.499352 \\
%%  5 &  13.195446 \\
%%  6 &  16.347210 \\
%%  7 &  19.904898 \\
%%  8 &  23.841461 \\
%%  9 &  28.139375 \\
%% 10 &  32.785980 
%% \end{array}
%% \]
%% \begin{array}{r|ccccccc}
%% s & 2 & 3 & 4 & 5 & 6 & 7 & 8 \\
%% \hline
%% \ka_s &
%% 11.1081 & 11.7318 & 14.0568 & 17.1853 & 20.9129 & 25.1578 & 29.8772 
%% \end{array}
% A more detailed description of $\ka_s$ is contained in
% Proposition~\ref{prop.LB}.
% A number of comments are in order.
As a corollary, by using
\be
0 \leq \frac{1}{\mmin} \leq 
\frac{1}{\ln 4}
\frac{\ln(s \, \mmin r_{\mmin})}{\mmin} \leq 
\frac{1}{\ln 4}
\phi(\vr)
\ef,
\ee
from (\ref{eq.3654635}) we can deduce bounds in a weaker but
functionally simpler form
\[
\frac{1}{\ka_s}
\phi(\vr) 
\leq \Phi_{\rm aver}(\vr) \leq \Phi_{\rm max}(\vr)
% \leq \Phi(\vr) 
\leq 
\bigg( 2+ \frac{1}{\ln 4} \bigg)
\phi(\vr)
\ef,
\]
i.e.\ the whole fuctional dependence from the dictionary is captured
by the function $\phi$, up to a 
multiplicative factor depending only on $s$, and bounded by 
$\sim 13.6 \frac{\sqrt{s}}{\ln s}$.

%% Note that, from the asymptotics of $\ka_s$, the forementioned
%% multiplicative factor is no bigger than $8 \ln s + \mathcal{O}(1)$,
%% i.e.\ we have a bound of the form (\ref{eq.boukk}) for $\kappa=
%% \sqrt{8 \ln s} + \mathcal{O}(1)$, if $\phi(\vr)$ is properly rescaled.

The reader may be wondering why the function $\phi(\vr)$ is the `good
one' to capture the behaviour of the complexity. We cannot give a full
intuition of this feature without entering in the details of the
calculations. Nonetheless, we can remark that $\phi(\vr)$ is defined
as a maximum over $m$, of a function of $m$ and $r_m$ alone. As a
result, from the statement of the theorem we deduce the following
striking property. Let $D$ be a dictionary for the generic MPMP, with
content $\vr=\{r_m\}_{m \geq 2}$.
%, realising the worst case for this choice of $\vr$.  
Let $m^*$ be any argmax of the function entering the definition of
$\phi$, for such a $\vr$, and let $D'$ be the dictionary corresponding
to the restriction of $D$ only to the words with length $m^*$, plus a
single word of length $\mmin$ (if $\mmin \neq m^*$).  As the functions
$\phi(\vr(\cdot))$ and $1/\mmin$ evaluate to the same values for $D$
and $D'$, we have thus determined that $\Phi(D)/\Phi(D') \leq
\kappa^{2}$, i.e.\ the pruned dictionary has a complexity not sensibly
smaller than the original one, despite the fact that the space of
possible $D$'s is not compact, and, even worse, $D$ may consist of an
infinite dictionary.
%% This, in turns, has implications both on $\Phi_{\rm aver}(\vr)$ and on
%% $\Phi_{\rm max}(\vr)$, when compared to the quantities associated to
%% $\vr'=\vr(D')$.

This may sound surprising, as one could have expected that, when $m^*$
is large, the function $\frac{1}{m} \ln (s\, m \, r_m)$ may well be
almost flat in an interval whose width scales with $m^*$ (e.g.,
between $\frac{1}{2} m^*$ and $2m^*$), and, as a result, $D$ has many
more words than $D'$.
% (up to exponentially-many more, as a function of $m^*$). 
One may have expected that the worst-case ratio $\Phi(D)/\Phi(D')$
showed a non-trivial functional dependence from $m^*$ under such
circumstances, e.g.\ a factor related to the width of such window, but
this is not the case, although, yet again, we cannot explain why
without entering the detailed analysis of the proof.

Let us make one last comment on the structure of the theorem.  Our
goal is the determination of the functional dependence of $\Phi(D)$
\emph{up to a finite multiplicative constant.}  From the forementioned
trivial bounds $\frac{1}{\mmin} \ln(s) \leq \Phi(D) \leq \ln(s)$, we see
that this goal only makes sense when the ratio between the two trivial
bounds is large, and there is no point in analysing dictionaries in which
$\mmin$ is smaller than this constant.

As a further corollary, there is no point in considering dictionaries
which do not have complexity $o(1)$ w.r.t.\ some size parameter.  It
is not hard to see that such dictionaries exist, even with simple
arguments unrelated to our analysis.\footnote{For example, in
  searching the word
${\tt aa \cdots a}$ of length $m$, in a two-symbol alphabet
$\Sigma=\{{\tt a},{\tt b}\}$, by the Boyer--Moore algorithm, we get
the asymptotic complexity $\frac{2}{m-1} +\mathcal{O}(2^{-m})$, which is
an upper bound to the exact complexity of this dictionary.}

The fact that `interesting' dictionaries have all words `sufficiently
long' leads to a \emph{leitmotif} of our analysis (which also appears
transversally in most of the mentioned literature). In the
construction of the bounds we are led to analyse complicated functions
$f(m)$, e.g.\ the solution of a transcendental equation. In order to
work out our results, we will just need that we can estimate upper
and/or lower bounds to this function, which are `effective' in a
regime of $m$ large, even if they are quite poor for $m$ too small.
As we will see, most functions $f(m)$ occurring in this problem will
allow for a perturbative expansion in powers of $(\ln m)/m$.

\subsection{Previous results}
\label{ssec.prevres}
%-------------------------------------------------------

\noindent
{\bf Single-word dictionaries.}\rule{10pt}{0pt}The 
statement of Theorem \ref{thm1}
restricted to single-word dictionaries is related to the result of the
seminal paper of Yao \cite{yao}. In this case, in $\phi$ there is no
need to take a $\max$, and $r_{m'}=1$ for $m=m'$ and $0$
otherwise. While our theorem gives that, for such a $\vr$, in worst
case among words of length $m$,
\[
\frac{1}{\kappa_s} 
\left( \frac{\ln(s\, m) +\frac{1}{s}}{m} \right)
\leq \Phi(\vr) 
\leq 2 
\left( \frac{\ln(s\, m) +\frac{1}{s}}{m} \right)
\ef,
\]
Yao determines
% more precisely 
that for almost all words of length $m$, and treating $s$ as a constant,
the complexity is $\Theta( \frac{\ln(m)}{m} )$.
%\footnote{There is a
% $\ln s$ relative factor among the two definitions of complexities.}
In order to establish this result, Yao introduces a notion of
\emph{certificate}. The set $C \subseteq \{1,\ldots,n\}$ is a
certificate for the pair $(D,\bxi)$ if the knowledge of 
$\{\xi_j\}_{j \in C}$ completely determines the output of the problem
(in fact, Yao defines certificates only for single-word dictionaries,
but the generalised notion is evinced in a natural way). Call
$\cC(D,\bxi)$ the set of certificates, and define
\be
\label{eq.phicert}
\Phi_{\rm cert}(D,\bxi):= 
\frac{\ln s}{n} \min_{C \in \cC(D,\bxi)} |C|
\ef,
\ee
together with the average quantities
\be
\Phi_{\rm cert}(D) := 
\eee_{\bxi} \big( \Phi_{\rm cert}(D,\bxi) \big)
\ef;
\ee
and
\be
\Phi_{\rm cert}(\vr) := 
\eee_{D\,:\vr(D)=\vr} \big( \Phi_{\rm cert}(D) \big)
\ef.
\ee
This is another natural notion of complexity for MPMP, that we
shall call the \emph{certificate complexity} of the instance. It is
the smallest possible complexity of the given instance,
or, in other words, it is the smallest complexity among
all possible runs of the probabilistic algorithm, that reads the
characters of the text one by one in a random order, up to when a
certificate is reached.

% Some reflection shows that 
This notion is a lower bound to the `true' complexity of
the instance,
% In formulas,
%
%\eee_{\bxi} 
% $\Phi_{\rm cert}(D,\bxi) \leq \Phi(D,\bxi)$, 
just because any
algorithm (and any run of a non-deterministic algorithm) may halt only
when a certificate is reached.
% , thus, from the minimum in the definition
% (\ref{eq.phicert}), it shall read at least 
Intuitively, reaching the value $\Phi_{\rm cert}(D,\xi)$ supposes not
only (or better, not necessarily) the optimality of the algorithm, but
also an infinite amount of `luck' in the choice of the characters to
read, given the hidden text $\bxi$, from which the inequality would
follow.

% As, by definition, $\Phi(D) \leq \Phi_{\cA}(D)$ for any given algorithm, 
Thus the analysis of $\Phi_{\rm cert}$ provides a natural proof
strategy for what concerns the estimate of a lower bound, that Yao
succeeds in pursuing for the single-word case, and that, in this
paper, we adapt to the case of general dictionaries.

As a side remark, it is natural to conjecture, although we suppose
hard to prove in full generality, that $\Phi_{\rm cert}(D) < \Phi(D)$,
i.e.\ that the necessity of `infinite luck' is intrinsic to the
definition and cannot be compensated by the optimisation of the
algorithm. This can be established rigorously, for example, for
$\Sigma= \{ {\tt a}, {\tt b}\}$ and dictionary $D=\{ {\tt ab}\}$, for
which one can determine that the Boyer--Moore algorithm is optimal,
and has information complexity $\Phi=5/6$, while the certificate
complexity is
$\Phi_{\rm cert}=13/16$, which is smaller by a tiny amount, roughly
$2\%$ (we do not illustrate this interesting fact here, as it would be
a \emph{d\'etour} w.r.t.\ our main aim). The gap between the two
complexities is expected to become increasingly smaller as the
dictionary gets larger, this making the forementioned conjecture more
challenging.

\bigskip
\noindent
{\bf Uniform dictionaries.}\rule{10pt}{0pt}Let us say that a dictionary $D$ is \emph{uniform} if all the words
have the same length. Fredriksson and Grabowski \cite{FG} describe a
MPMP algorithm (in the following, the \emph{FG algorithm}) adapted to
deal with uniform dictionaries, and analyse the resulting upper bound.
This algorithm is possibly never optimal (and known to be non-optimal
on certain simple dictionaries), but the loss is by a relatively small
factor, and, in many applications, is compensated by a great
simplicity both in programming and in the statistical analysis.  Based
on previous results of Fredriksson and Navarro \cite{FN}, that
purportedly estimated lower bounds for uniform dictionaries, the
authors claimed their algorithm to be optimal up to multiplicative
factor, in its domain of contents $\vr$, and provided a complexity
compatible with the corresponding specialisation of our theorem above.
%% Our initial aim was to extend the algorithm in \cite{FG} to arbitrary
%% dictionaries, and adapt the analysis of both the upper- and
%% lower-bounds. However, as we have become aware at a later stage, 
However, the results of \cite{FN} are invalidated by a major
flaw\footnote{See the annotation at {\tt
    https://www.dcc.uchile.cl/}$\sim${\tt gnavarro/fixes/tcs04.html},
  based on a personal communication of Ralf Stiebe to G.~Navarro.}
(and thus the optimality result in \cite{FG} remained unproven before
the present paper).  For this reason, our lower-bound estimate, for
dictionaries of arbitrary content, is completely different from the
one in \cite{FN}, while it is in fact mostly an adaptation of the
original strategy of Yao~\cite{yao} for single-word dictionaries. 

This lower bound is presented in Section 4, while the adaptation of
the
% Fredriksson and Grabowski 
FG algorithm, and the resulting 
upper bound, are outlined in Sections 2 and 3, respectively.
\section{The Fredriksson--Grabowski algorithm}
As we mentioned, ideally we want an algorithm with a degree of
flexibility, such that, in the interesting cases $\mmin \gg 1$, a
fraction $\ll 1$ of the text is accessed, i.e.\ some `big jumps' are
performed.  The idea of Fredriksson and Grabowski \cite{FG} is to
encode this flexibility in a single parameter $q$ associated to a
\emph{dilution rate} of the words in the dictionary.

For a fixed integer $q$, a text $\bxi$ and a dictionary 
$D=\{w_j\}_{1 \leq j \leq k}$, define the \emph{diluted text}
$\bxi^{(q)}=(\xi_{q}, \xi_{2q}, \xi_{3q}, \ldots)$ and the
\emph{diluted dictionary}
$D^{(q)} = \{w_{j,p} \}_{1 \leq j \leq k,\; 0 \leq p < q}$.  
For a word $w_j$ of length $m$, call $w_{j}^1$, \ldots, $w_{j}^m$ its
$m$ characters.
If
$|w_j|=m_j$, $w_{j,p}$ is defined as 
\[
w_{j,p} =
w_j^{m_j-p-h q}
\cdots 
w_j^{m_j-p-q} 
w_j^{m_j-p} 
\qquad
h=\lfloor (m_j-p-1)/q\rfloor
% 1\leq m_j-p-h q \leq q
\ef.
\]
For example, if $w_1=\textrm{\tt thepattern}$ and $q=3$, the three
words $w_{1,0}$, $w_{1,1}$ and $w_{1,2}$ are obtained by the sieve
construction

\begin{center}
\begin{tabular}{l|l}
{\tt t~h~e~p~a~t~t~e~r~n} & \\
\hline
{\tt t~.~.~p~.~.~t~.~.~n} & {\tt tptn} \\
{\tt .~.~e~.~.~t~.~.~r~.} & {\tt etr} \\
{\tt .~h~.~.~a~.~.~e~.~.} & {\tt hae} 
\end{tabular}
\end{center}

\noindent
Call $[m/q]_p := \lfloor \frac{m-p-1}{q} \rfloor +1$. Then
$|w_{j,p}|=[|w_{j}|/q]_p$.

% If $|w_j|<q$, the resulting empty words will be discarded.

Let us denote by $\FG(q)$ the algorithm of parameter $q$, and by
$\Phi_{\FG(q)}(D)$ the upper bound to $\Phi(D)$ obtained by analysing
this algorithm.  The FG algorithm of parameter $q$, in its simplest
version, consists in reading the characters of $\bxi^{(q)}$ in
sequence, and searching for words in the dictionary $D^{(q)}$.  Once
an occurrence has been found, it reads the missing $m\frac{q-1}{q} +
\mathcal{O}(1)$ characters in order to verify if the original word was
present. This idea can be improved at various extents. For example,
once we have agreed to solve the MPMP $(D^{(q)}, \bxi^{(q)})$, nothing
prevents from using a `good' algorithm for this, and possibly, just
recursively the FG algorithm, instead of switching to the na\"ive
read-all algorithm. Furthermore, instead of just filling the whole
gap, we can just proceed to read the characters of the gap one by one,
and stop the procedure if a contradiction with the candidate word is
found (for future reference, let us call this \emph{sharp gap
  filling}). However it turns out that, at the aim of determining
$\Phi(D)$ up to multiplicative constants, these improvements are not
crucial.  See Figure \ref{fig.exFG} for an example.

\begin{figure}[tb]
\setlength{\unitlength}{12pt}
\begin{center}
\begin{picture}(25,5)(0,-2)
\put(0,1.5){$D=\{w\}=\{ \textrm{\tt thepattern} \}$\;;\quad $q=3$; \quad
$D^{(q)}= \{ \textrm{\tt tptn}, \textrm{\tt hae}, \textrm{\tt etr}
  \}$}
\put(0,1){\line(1,0){25}}
\put(0,0){$\xi=$\; {\tt .~.~\underline{x}~.~.~\underline{h}~.~.~\underline{a}~s~t~\underline{e}~r~n~\underline{t}~.~a~\underline{r}~n~.~\underline{r}~.~.~}}
\put(0,-1){\phantom{$\xi=$\; {\tt .~.~\underline{x}~.~}}{\tt \gra{t}~\underline{h}~\gra{e}~\gra{p}~\underline{a}~\red{s}~t~\underline{e}~r~n}}
\put(0,-2){\phantom{$\xi=$\; {\tt
      .~.~\underline{x}~.~.~\underline{h}~.~.~\underline{a}~}}{\tt \gra{t~h}~\underline{e}~\gra{p~a}~\underline{t}~\gra{t}~\red{a}~\underline{r}~n}}
\end{picture}
\end{center}
\caption{A typical realisation of the Fredriksson--Grabowski
  algorithm. Dotted entries will not be read. In its basic
  implementation, analysed here, it is the {\tt a} at the far right
  that excludes the decimated pattern {\tt etr}. A more efficient
  variant, but more complicated to analyse, would find that the
  pattern {\tt etr} is excluded by the $\xi_i$'s already read at the
  previous step.\label{fig.exFG}}
\end{figure}

In the (non-improved) $\FG(q)$ algorithm, we read at least a fraction
$1/q$ of the text, and, as it will turn out, for the optimal value of
$q$ the total fraction is slightly above $1/q$. If a dictionary $D$
has its shortest word of length $\mmin$, the complexity of any
algorithm is at least $1/\mmin$, so, under the \emph{ansatz} above,
there is no point in searching among values $q > \mmin$, and in
particular there are no empty words in $D^{(q)}$.

The exact value of $\Phi_{\FG(q)}(D)$ is hardly expressed by a closed
formula, because we may be forced to fill the same gap by more than
one occurrence of diluted words. These characters shall count only
once in the complexity, but the interplay of different words in the
dictionaries is complicated to analyse.  So, we will content ourselves
with the upper bound to this complexity obtained from the union bound
(Boole's inequality\footnote{I.e., the fact 
$\mathrm{prob}(\mathcal{E}_1 \vee \cdots \vee \mathcal{E}_k)
\leq \mathrm{prob}(\mathcal{E}_1) + \cdots +
\mathrm{prob}(\mathcal{E}_k)$.}) 
on these occurrences:
\begin{proposition}
\be
\label{eq.423424}
\Phi_{\FG(q)}(D)
\leq
\frac{\ln s}{q}
\bigg(
1 + 
\sum_m r_m 
\sum_{p=0}^{q-1} s^{-[m/q]_p}
(m-[m/q]_p)
%{\lfloor \frac{m+p}{q} \rfloor}
\bigg) \ef.
\ee
\end{proposition}
{\it Proof.} The formula above is deduced from an easy reflection on
the mechanisms of the algorithm. We read one in $q$ characters, to
solve the diluted problem with the trivial read-all algorithm (from
which the $1/q$ factor).\footnote{As we said, for the \emph{ordinary}
  complexity paradigm, this problem is \emph{not} trivial, as it
  requires the use, for example, of the Aho--Corasick
  automaton. However, for what concerns the number of accesses to the
  text, reading all the characters is just the worst possible
  strategy, and thus a trivial upper bound.}  Then, at each position
of the diluted string, if we find a candidate subpattern $w_{j,p}$, we
read the missing $|w_j| - |w_{j,p}|$ characters (from which the
$(m-[m/q]_p)$ factor). This gives an overcounting (by union bound)
when we find more than one subpattern at the same position, or at a
short distance along the diluted text, but it makes the probabilities
of finding subpatterns independent, and thus just given by the obvious
factor $s^{-|w_{j,p}|}$, this being an important simplification to the
formula, relevant in the following calculations. \qed

Note how, just by the use of the read-all algorithm on the diluted
subproblem, and the union bound on the occurrences in this subproblem,
we have determined an upper bound on $\Phi_{\FG(q)}(D)$ that depends
on $D$ only through its content $\vr(D)$.  This upper bound on the
complexities of all variants of the FG algorithm discussed above is in
fact the \emph{exact} complexity of the most na\"ive version of the
algorithm, the one in which, if more than one word of $D^{(q)}$ is
found at a given diluted position, the missing characters are accessed
multiple times, and these multiple accesses are counted with their
multiplicities.\footnote{Note that, for such a paradigm, in principle
  the trivial upper bound $(\ln s) \Phi_{\FG(q)}(D) \leq n$ may be
  violated, and it will be our care to choose appropriate values of
  $q$ such that the bound is tight.}

If we perform the `sharp gap filling' enhancement discussed above,
i.e.\ we stop to read characters in the verification of a given
candidate sub-pattern whenever we obtain a contradiction, we can
replace the factor $m':= m-[m/q]_p$ in (\ref{eq.423424}), i.e.\ the
maximal number of read characters for the diluted pattern, by the
\emph{expectation} of read characters, which is
\be
\smfrac{s-1}{s}
\left(1+
% 2 s^{-1}+3 s^{-2}+\cdots + (m'-1) s^{-(m'-2)} 
\smfrac{2}{s}
+
\smfrac{3}{s^2}
+\cdots + 
\smfrac{m'-1}{s^{m'-2}}
\right)
+
\smfrac{m'}{s^{m'-1}}
% m' s^{-(m'-1)} 
=
\smfrac{s}{s-1} (1-s^{-m'})
\ee
getting the improved expression
\begin{proposition}
\be
\label{eq.764656}
\Phi_{\FG(q)}(D)
\leq
\frac{\ln s}{q}
\bigg(
1 + 
\frac{s}{s-1} 
\sum_m r_m 
\sum_{p=0}^{q-1}
\big( s^{-[m/q]_p} - s^{-m} \big)
\bigg)
\ef.
\ee
\end{proposition}
%

%% In the following section, through a sequence of
%% further bounds, we give an estimate of this quantity in terms of the
%% function $\phi(\vr)$.

%%%%%%%%%%%%%%%%%%%%%%%%%%%%%%%%%%%%%%%%%%%%%%%%%%%%%%%
\section{The upper bound}
%
% Assume $\mmin(D) \geq m_0$. 

We shall now analyse the expression on the RHS of the inequality
(\ref{eq.423424}) -- or better, the inequality (\ref{eq.764656})
improved by the sharp gap filling -- in order to produce a more
compact expression for an upper bound to the complexity, and, in
agreement with Theorem~\ref{thm1}, prove the following
\begin{proposition}
With $\phi(\vr)$ as in equation (\ref{eq.phir}), we have
\label{prop.UB}
\be
\label{eq.ub65454}
\Phi_{\rm max}(\vr) \leq
% \Phi_{\FG(\ln s/(2\phi(\vr)))}(\vr)
% \leq
2 \phi(\vr) 
% \left(
+ 
% \frac{\ln s}{s-1}
% \frac{m_0}{m_0-1}
\frac{1}{s\,\mmin}
%% \frac{s^2 \ln s}{s-1}
%% \frac{1}{2m_0-4}
% \right)
\ef.
\ee
\end{proposition}
\noindent
As we have $\Phi(D) \leq \Phi_{\FG(q)}(D)$ for any value of $q$, we
have in particular $\Phi(D) \leq \Phi_{\FG(q(\vr(D)))}(D)$ for any
integer-valued function $q(\vr)$, the best choice being the value of
$q$ minimising the expression on the RHS of (\ref{eq.764656}), for the
given $\vr$.  We shall provide a function $q(\vr)$ which, although not
being exactly the minimum, is sufficiently close to this value that
our upper and lower bounds, as functions of $\vr$, match up to a
finite multiplicative constant. However, this condition can only be
verified
\emph{a posteriori}, once that a lower bound, of functional form
analogous to the RHS of (\ref{eq.ub65454}), is established, so in this
section we shall content ourselves with the analysis of the resulting
expression, leaving our choice of $q(\vr)$ as an \emph{ansatz}.

\vspace{2mm}
\noindent
{\it Proof of Proposition \ref{prop.UB}.}
Let us start from the improved expression
(\ref{eq.764656}). Let us drop the summands $-s^{-m}$, and get the
slightly larger bound
\be
\label{eq.764656bis}
\Phi_{\FG(q)}(D)
\leq
\frac{\ln s}{q}
+
\frac{s \ln s}{s-1} 
\sum_m r_m 
\frac{1}{q}
\sum_{p=0}^{q-1}
s^{-[m/q]_p}
\ef.
\ee
Let us concentrate on the inner-most expression
\be
f(s;m,q):=
\frac{1}{q}
\sum_{p=0}^{q-1}
s^{-[m/q]_p}
\ef.
\ee
Writing $m=aq+b$ with $0 \leq b < q$, this just reads
\be
f(s;aq+b,q):=
s^{-a}
\frac{b s^{-1} + q-b}{q}
\ef.
\ee
This function can be analytically extended, in terms of
\be
g(s,x):=s^x \big( x s^{-1} + 1-x \big)
\ee
because
\be
f(s;m,q) = s^{-\frac{m}{q}} g(s,b/q)
\ef.
\ee
For $s>0$, the function $g(s,x)$ is concave on $[0,1]$, its only
maximum is at $x^*(s)=1+\frac{1}{s-1}-\frac{1}{\ln s}$, where one has
$g^*(s)=g(s,x^*(s))=\frac{s-1}{\ln s} \exp\big(\frac{\ln s}{s-1}-1\big)$.
In particular, for $s \geq 2$ we have
$g^*(s) \leq \frac{2}{e} \frac{s-1}{\ln s}$.
Substituting into (\ref{eq.764656bis}) gives
\be
\label{eq.764656ter}
\Phi_{\FG(q)}(D)
\leq
\frac{\ln s}{q}
+
\frac{2s}{e}
\sum_m r_m 
s^{-\frac{m}{q}}
\ef.
\ee
The change of variables $\rho=\frac{\ln s}{q}$ gives
\be
\label{eq.764656b}
\begin{split}
&\Phi_{\FG(q)}(D)
\leq
\rho
+
\frac{2s}{e}
\sum_m r_m 
e^{-m \rho}
=
\rho
+
% \frac{s^2 \ln s}{s-1}
\frac{2}{es}
\sum_{m \geq \mmin} 
\frac{e^{-m \big( \rho - \frac{\ln (s^2 m^2 r_m)}{m} \big)}}{m^2}
\ef.
\end{split}
\ee
Now choose the candidate value $\rho=2 \phi(\vr)$, that is,
$q(\vr)=\ln s/(2 \phi(\vr))$, with $\phi(\vr)$ as in equation
(\ref{eq.phir}).\footnote{Up to a round-off due to $q$ being an
  integer. We neglect here the (easy) corrections coming from this
  feature.}
We have $\rho \geq \frac{2 \ln (s m r_m)}{m} \geq \frac{\ln (s^2 m^2 r_m)}{m}$ 
for all $m$, so that
% \footnote{We implicitly use $\ln(s)/(s-1) < 1$ for all integers $s
%  \geq 2$.}
%
\bes
\begin{split}
&
\sum_{m \geq \mmin} 
\frac{e^{-m \big( \rho - \frac{\ln (s^2 m^2 r_m)}{m} \big)}}{m^2}
\leq
\sum_{m \geq \mmin} 
\frac{1}{m^2}
% \leq
% \frac{1 + \frac{1}{2\mmin}+ \frac{1}{6\mmin^2}}{\mmin}
\leq
% \frac{31}{24}
\frac{\pi^2-6}{3}
\frac{1}{\mmin}
%% <
%% \frac{1}{\ln 2}
%% \frac{1}{\mmin}
\ef,
%% \\
%% &\quad
%% =\frac{1}{\mmin-1}
%% \leq
%% \frac{\ln \mmin}{\mmin}
%% \frac{1}{2m_0-4}
%% % \frac{m_0}{2(m_0-1)(m_0-2) \ln m_0}
%% \leq
%% \phi(\vr)
%% \frac{1}{2m_0-4}
%% % \frac{m_0}{2(m_0-1)(m_0-2) \ln m_0}
\end{split}
\ees
(where we used the result of Appendix~\ref{app.546756}).
As $\frac{2}{e}\frac{\pi^2-6}{3}<1$, we can write
\be
\label{eq.ub65454xx}
% \Phi_{\rm max}(\vr) \leq
\Phi_{\FG(\ln s/(2\phi(\vr)))}(\vr)
\leq
2 \phi(\vr) 
+
\frac{1}{s\, \mmin}
\ef.
\ee
This allows us to conclude.
\hfill $\square$

%% \[
%% %\be
%% %\label{eq.764656b}
%% \Phi_{\FG(q)}(D)
%% \leq
%% \rho
%% +
%% \frac{s^2 \ln s}{s-1}
%% \sum_m r_m 
%% e^{-m \rho}
%% \leq
%% \rho
%% +
%% \frac{s^2 \ln s}{s-1}
%% \sum_{m \geq \mmin} 
%% \frac{e^{-m \big( \rho - \frac{\ln (r_m m^3)}{m} \big)}}{m(m-1)(m-2)}
%% \ef.
%% \]
%% %\ee
%% Now choose the candidate value $\rho=3 \phi(\vr)$.\footnote{Up to
%%   round-off due to $q$ being an integer. We neglect this easy
%%   term.} We have $\rho \geq \frac{\ln (r_m m^3)}{m}$ for all $m$, so
%% that 
%% %
%% \bes
%% \begin{split}
%% &
%% \sum_{m \geq \mmin} 
%% \frac{e^{-m \big( \rho - \frac{\ln (r_m m^3)}{m} \big)}}{m(m-1)(m-2)}
%% \leq
%% \sum_{m \geq \mmin} 
%% \frac{1}{m(m-1)(m-2)}
%% \\
%% &\quad
%% =\frac{1}{2(\mmin-1)(\mmin-2)}
%% \leq
%% \frac{\ln \mmin}{\mmin}
%% \frac{1}{2m_0-4}
%% % \frac{m_0}{2(m_0-1)(m_0-2) \ln m_0}
%% \leq
%% \phi(\vr)
%% \frac{1}{2m_0-4}
%% % \frac{m_0}{2(m_0-1)(m_0-2) \ln m_0}
%% \end{split}
%% \ees
%% (provided that $m_0 \geq 4$, so that $\ln m_0 \geq m_0/(m_0-1)$). Thus
%% we have found
%% \be
%% \Phi(\vr) \leq
%% \Phi_{\FG(3\phi(\vr))}(\vr)
%% \leq
%% \phi(\vr) \left(
%% 3 + 
%% \frac{s^2 \ln s}{s-1}
%% \frac{1}{2m_0-4}
%% \right)
%% \ef.
%% \ee

%%%%%%%%%%%%%%%%%%%%%%%%%%%%%%%%%%%%%%%%%%%%%%%%%%%%%%%
\section{The lower bound}
\label{sec.lb}

In this section we establish a lower bound on the complexity, averaged
over all dictionaries of content $\vr$, of the form

\begin{proposition}
\label{prop.LB}
For $s \geq 2$, define 
$\ka_s = 5\frac{\sqrt{s}}{\ln s}$.  In the MPMP on an
alphabet of size $s$,
\be
\Phi_{\rm aver}(\vr)
\geq
\Phi_{\rm cert}(\vr)
\geq
\frac{1}{\ka_s} 
\left( \phi(\vr) + \frac{1}{2s\, \mmin} \right)
\ef.
\ee
%% The value of $\ka_s$ is the solution of 
%% \begin{align}
%% \ka_s 
%% &=
%% \min_{\tau} \left[ \; \max_{Q} \left[ \; \frac{1}{2 \tau
%%       (1-\exp(-F
%% %(Q,\tau,s)
%% ))}
%% \; \right] \right]
%% \ef;
%% \\
%% F
%% %(Q,\tau,s)
%% &=
%% \frac{ e^{Q (1-\tau) - 2 a \tau}  a^3}{Q(Q+a)^2}
%% -
%% \frac{ e^{-Q (1-\tau) + 2 a \tau} Q(Q+a)^2}{a^3}
%% - \frac{Q(Q+a)}{a}
%% \ef;
%% \\
%% a&= \ln s/\tau
%% \ef.
%% \end{align}
%%  smallest real value such that, posing
%% \begin{align}
%% c &= \frac{ e^{Q (1-1/\ka)}  a^3}{s^2 Q(Q+a)^2}
%% \ef;
%% &
%% a&= \ka \ln s
%% \ef;
%% \end{align}
%% the quantity $c-c^{-1}- \ln 2 - \frac{Q(Q+a)}{a}$ is positive for all
%% $Q>0$.
\end{proposition}

%-------------------------------------------------------
\subsection{Reduction to a finite text}

In evaluating a lower bound on the information complexity of MPMP, it
is legitimate to replace the exact expression by simpler quantities
which bound it from below. The goal is to reach a concretely
evaluable expression, without losing the leading functional dependence
from $D$ (up to multiplicative constants), or, better, the dependence
from~$\vr(D)$.

The output of a MPMP is a list of pairs $S(D,\bxi) = \{(\ell,h)\}$, with
$1 \leq h \leq k$ and $1 \leq \ell \leq n-|w_h|+1$, such that $w_h$
occurs at position $\ell$ in the text. Call 
$X_0=\{ (\ell,h) \,|\, 1 \leq h \leq k \,;\, 1 \leq \ell \leq n-|w_h|+1 \}$ 
the list of possible pairs described above.  If we specify in advance
a subset $X \subseteq X_0$, and decide that we aim at finding all the
occurrences within this subset, we have a reduced problem with
complexity, say, $\Phi(D|X)$ (while the complexity of the original
problem is $\Phi(D) \equiv \Phi(D|X_0)$).  The problem associated to
the subset $X$ is evidently simpler than the original one, at least
within our notion of complexity (in terms of text characters to be
read): a solution $S(D,\bxi)$ of the full problem provides a solution
of the reduced one, just by taking the intersection $S(D,\bxi) \cap
X$, and this process does not involve further accesses to the text,
thus we have $\Phi(D|X) \leq \Phi(D)$ for all $X \subseteq X_0$.  A
similar argument applies to the certificate complexity.  Thus, this
observation provides a useful strategy for producing lower bounds for
both versions of the complexity.

As it was the case for the upper bound and the parameter $\rho$, we
need a free real parameter in order to capture the functional
dependence from $D$. In analogy to Yao procedure, a good choice is an
integer $L$, which determines a set $X=X(L)$ as follows:
\be
\label{eq.defXL}
X(L) = 
\left\{ (\ell,h) \,\bigg|\,
\left\lfloor \frac{\ell+1}{L} \right\rfloor 
= 
\left\lfloor \frac{\ell+|w_h|}{L} \right\rfloor
\right\}
%% \{ (\ell,h) \,|\,
%% \lfloor (\ell+1)/L \rfloor = \lfloor (\ell+|w_h|)/L \rfloor
%% \}
\ef.
\ee
% $(\ell,h) \in X$ if
% $\lfloor (\ell+1)/L \rfloor = \lfloor (\ell+w_h)/L \rfloor$, i.e., 
In simple words, the text, originally of length $n$, is divided into
blocks of size $L$ (with the possible exception of one last block of
size $<L$), and we search for occurrences contained within one
block. As a result of this bound, we have factorised the problem on
the blocks, and the calculation of (the bound on) the complexity rate,
i.e.\ the limit of $1/n$ times the complexity, for $n \to \infty$, is
given by $1/L$ times the whole average complexity of single blocks,
again both for ordinary and certificate complexities.
\begin{align}
\label{eq.blockbound}
\Phi(D) 
& \geq 
% \frac{1}{L}\,
\eee_{\bxi \in \Sigma^L} \Phi(D,\bxi)
\ef;
&
\Phi_{\rm cert}(D) 
& \geq 
% \frac{1}{L}\,
\eee_{\bxi \in \Sigma^L} \Phi_{\rm cert}(D,\bxi)
\ef;
&&
\forall \ L
\ef.
\end{align}
This solves the issue of performing the limit $n \to \infty$, and
reduces to a problem which, for a fixed dictionary, is of finite size,
namely $L$ (although, in fact, the optimal value of $L$ as a function
of $D$ is unbounded).  From this moment on, we can thus concentrate on
a single block.
%% As well as, in the previous section, we needed the integer parameter
%% $q$ to be not above the relevant scale $m$ of the dictionary (a
%% value $q \sim m/\ln m$ turned out to be optimal), here
% The value of $L$ has to be suitably chosen. 

It is rather clear that we need $L$ to be not below the scale $m^*(D)$
of the most relevant word-length for the given dictionary, otherwise
we would just miss the totality of the occurrences which give the
leading term of the complexity.  Surprisingly, it will turn out that
the optimal value of $L$ is barely above $m^*$.

%-------------------------------------------------------
\subsection{The lattice of possible algorithms}

Under our complexity paradigm, and for a fixed text $\bxi$ of length
$L$, we can visualise any algorithm as a recipe for performing a
directed walk on the lattice of subsets $I$ of $\{1,\ldots,L\}$.

%% A general (deterministic) algorithm can be represented by a decision
%% tree. The tree has internal nodes, where some operations have to be
%% performed, and leaves, at which a certificate has been reached, and
%% the solution is returned.

%% In our specific problem, of MPMP on a text of length $L$, because of
%% our complexity paradigm, we can suppose that the decision tree is
%% graduated by the number of accesses to the text, so that the height in
%% the tree at which a certificate is reached coincides with the complexity.

%% We aim at understanding the behaviour of the \emph{optimal} algorithm,
%% as it determines directed walks on the decision tree. As the notion of
%% optimal algorithm is somewhat slippery, even at the risk of looking
%% verbose, we want to state clearly how an algorithm works.

\begin{figure}[tb]
\[
\includegraphics[width=\textwidth]{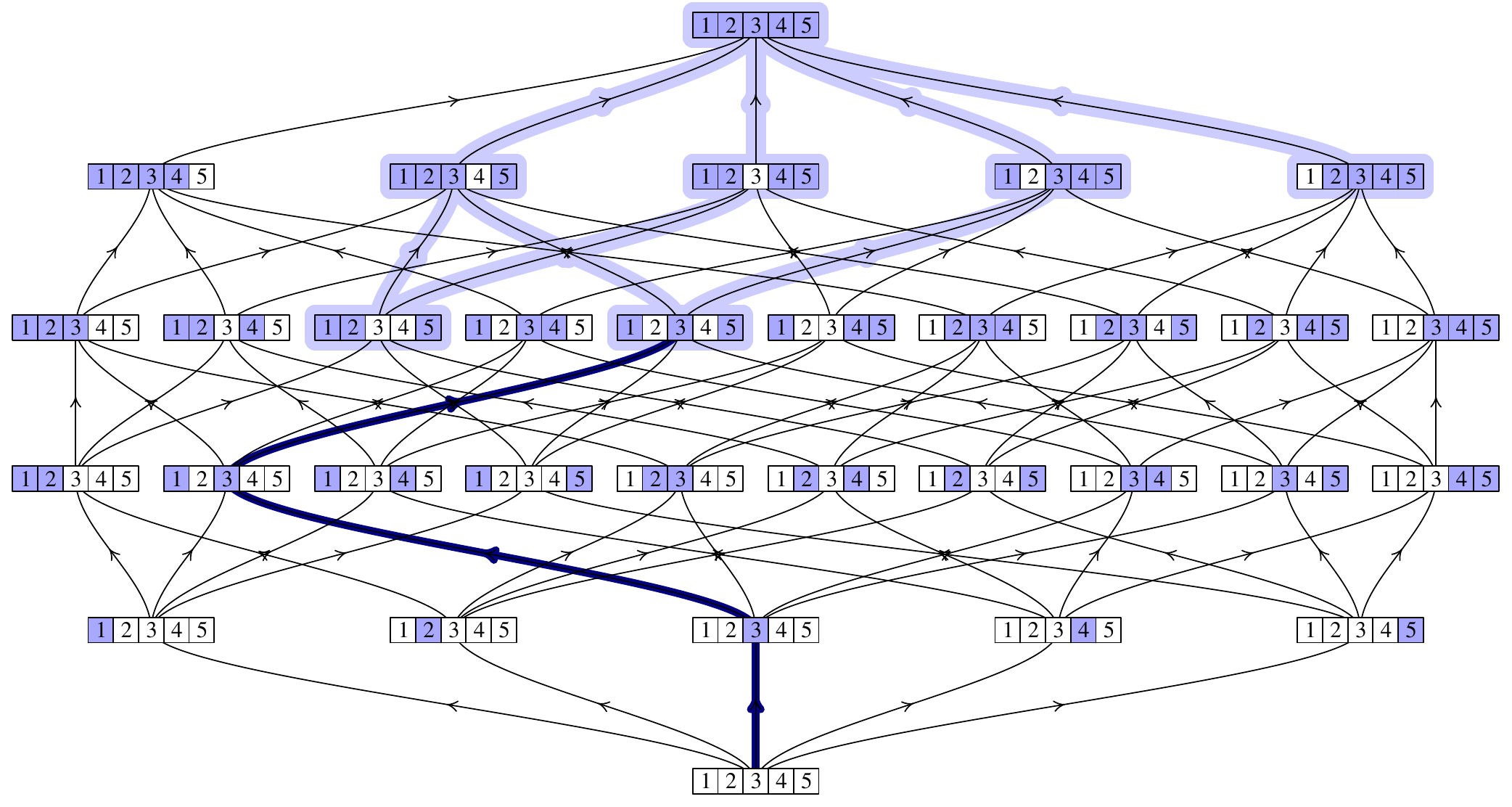}
\]
\caption{\label{fig.latt}The lattice of subsets of
  $\{1,\ldots,5\}$. The shadowing denotes an example of up-set,
  corresponding, for a certain pair $(D,\bxi)$, to certificate
  nodes. The thick path is a typical run of an algorithm, which halts
  whenever it reaches a shadowed node, in the example at $t=3$.}
\end{figure}

Any algorithm starts at the bottom of the lattice, i.e.\ the node
associated to the empty subset.  The algorithm evaluates a function of
$D$, in order to choose a position $1 \leq i_1 \leq L$ to access.
Then, it reads $\xi_{i_1}$, thus `moving' to the set $\{i_1\}$, and it
evaluates a function of $D$ and $\xi_{i_1}$, in order to choose a
position $i_2 \neq i_1$.  Then, it reads $\xi_{i_2}$, thus moving to
$\{i_1,i_2\}$, and chooses a position
$i_3 \neq i_1, i_2$ through a function of $D$ and
$\{\xi_{i_1},\xi_{i_2}\}$, and so on, until a certificate is reached
(see Figure \ref{fig.latt}).
We say that the algorithm, at a certain moment of its execution, is at
the node $I=\{i_1,\ldots,i_t\}$ of the lattice, if, so far, it has
accessed the text characters at positions $\{i_1,\ldots,i_t\}$ (in any
of the possible orders).  The nodes are associated to unordered sets
of positions, instead that of ordered lists, because there is no
reason why the function of $D$ and
$\{\xi_{i_1}, \ldots, \xi_{i_t} \}$ shall depend on the order in which
the positions $\{i_1,\ldots, i_t\}$ have been accessed: any such
dependence would only make the algorithm less performing.

A set $I$ may or may not be a \emph{certificate} for the given text
and dictionary, i.e.\ it may or not contain enough information to
determine the set $S(D,\bxi)$ of occurrences. Clearly, if $I$ is a
certificate and $J \supseteq I$, also $J$ is a certificate, so that,
for fixed $\bxi$ and $D$, the certificates are an up-set in the
lattice.  Following Yao, we call \emph{negative certificate} a set
certifying that $S(D,\bxi)=\varnothing$.

Let us call $\chi_{D,\bxi}(I)$ the boolean function valued 1 if the
knowledge of the text at positions $I$ does \emph{not} provide a
certificate, and 0 otherwise (i.e., we have $\chi_{D,\bxi}=1$ on the
bottom part of the lattice, and $\chi_{D,\bxi}=0$ on the top part).

If we neglect (for the moment) the walks which stop at a certificate,
and we consider the algorithm on a given text $\bxi$, we have an upper
bound on the number of distinct nodes at time $t$, which is
$\binom{L}{t}$.  Some walks may stop before time $t$ because they
reach a certificate.  In order to preserve the stochasticity of the
process, we can imagine that the algorithm continues (arbitrarily)
even when a certificate is reached.  As a result, at time $t$, the
walk determined by the algorithm may be at any set $I$ among the
possible $\binom{L}{t}$ ones, with a probability $P_{D,\bxi}(I)$ (we
consider probabilities because the optimal algorithm may be a
probabilistic one), these sets $I$ may or not be a certificate.

%%   If, as easily conjectured, there exists
%% one optimal algorithm which is deterministic, in fact
%% $P_{D,\bxi}(I)=1$ on a unique set $I$, and 0 elsewhere, but we would
%% not need this fact in the following.

The complexity $\Phi_{\cA}(D,\bxi)$ of the algorithm $\cA$ is, by
definition, the average depth of the nodes at which a certificate is
first reached, weighted with their branch probabilities.

If this average is evaluated for an algorithm which is \emph{optimal}
at given $D$ and $L$, and averaged over all text $\bxi$, it would
correspond exactly to $\Phi(D)$ for texts of length $L$. Thus, and by
using (\ref{eq.blockbound}), any lower bound constructed by analysing
this probabilistic process provides a lower bound to the complexity.
However, we have a very mild control on the probability distribution
$P_{D,\bxi}(I)$ associated to the optimal algorithm (because
constructing explicitly the optimal algorithm is a formidable
task). For this reason we rather concentrate on the simpler (and
smaller) quantity $\Phi_{\rm cert}(D)$, introduced in Section
\ref{ssec.prevres} (in fact, we will rather analyse the related
quantity $\Phi_{\rm cert}(\vr)=\eee_D\Phi_{\rm cert}(D)$), as we now
describe.

Say that $I$ is a set of size $t$, and introduce the quantity
\be
\label{eq.defFvr}
f_{\vr}(I) := \eee_D \Big(
1-\prod_\bxi \chi_{D,\bxi}(I) \Big)
\ef.
\ee
Informally, this is the average over all dictionaries of content
$\vr$, of the probability that the set $I$ is a certificate for some
text $\bxi$ (i.e., for some $t$-uple $(\xi_{i_1},\ldots, \xi_{i_t})
\in \Sigma^t$).

Call $1-p_t(\vr)$ the fraction of dictionaries of content $\vr$ which
have, for some text, some certificate of size at most $t$.  By the
union bound, this quantity is bounded from above by the average, over
the dictionaries, of the number of certificates of size at most $t$
(i.e., if there is a certificate $J$ of cardinality $t'$, which is
minimal w.r.t.\ inclusion, it gets weighted by $\binom{L-t'}{t-t'}$,
i.e.\ the number of sets $I' \supseteq J$ of cardinality $t$).  But
this latter quantity is, by definition, $\sum_I f_{\vr}(I)$, from
which we get
\be
\label{eq.58725764}
1-p_t(\vr)
\leq
\sum_I f_{\vr}(I)
\ef.
\ee
If, for the given text, a dictionary has no certificate of length at
most $t$, its complexity is at least $t+1$. If it has one or more such
certificates, its complexity is bounded from below at least by the
forementioned obvious bound $\lceil (L-\mmin+1)/\mmin \rceil$, which,
for large $L$, is essentially $L/\mmin$.  For this reason, and by
using (\ref{eq.blockbound}), we can conclude that
\be
\Phi_{\rm cert}(\vr) \geq 
% \frac{1}{\ln s} 
\ln s \; \max_t \Big(
\frac{t\, p_t}{L} + \frac{1-p_t}{\mmin} \Big)
\ef.
\label{eq.lb14452}
\ee
It will be our care to consider values of $t$ larger than $L/\mmin$
(otherwise we would make statements which are weaker than the trivial
bound!). In this case we can rewrite the expression above as
\be
\max_{t \geq \frac{L}{\mmin}} \Big(
\frac{(t+1)\, p_t}{L} + \frac{1-p_t}{\mmin} \Big)
=
\frac{1}{\mmin} 
+
\frac{1}{L}
\max_{t \geq \frac{L}{\mmin}} \Big(
p_t \Big(t+1-\frac{L}{\mmin} 
\Big)
\Big)
\ee
which is at sight a monotone increasing function of $t$ and $p_t$
(when considered as independent variables). For this reason, the bound
(\ref{eq.58725764}) provides an actual bound to the complexity, namely
\begin{align}
\Phi_{\rm cert}(\vr) &\geq 
% \frac{1}{\ln s} 
\ln s \; \max_{t  \geq L/\mmin} \Big(
\frac{
% (t+1)
t \, \tilde{p}_t}{L} + \frac{1-\tilde{p}_t}{\mmin} \Big)
\ef;
\label{eq.lb14452bis}
\\
\tilde{p}_t(\vr)
&=
1-
\sum_I f_{\vr}(I)
\ef.
\label{eq.lb14452bis2}
\end{align}
Of course, $t$ and $\tilde{p}_t$ are not independent variables, and in
fact $\tilde{p}_t$ is a monotonically \emph{decreasing} function of
$t$. This is what makes the optimisation problem non-trivial.  In the
remainder of this section we will estimate the quantity $\tilde{p}_t$,
at values $t$ and $L$, choosen as a function of $\vr$, which are near
to the optimal value, and thus, once plugged in (\ref{eq.lb14452bis}),
produce a bound of the desired functional form.

%-------------------------------------------------------
\subsection{Reduction to a problem on surjections}

Let us consider an algorithm, together with its lattice of executions.
Consider one given word $w \in D$, of length $m$, and one given set
$I=\{i_1,\ldots,i_t\}$ of size $t<m$, appearing as a node of depth $t$
in the lattice. Suppose that the algorithm has read the values
$\xi_{i_1}$, \ldots, $\xi_{i_t}$ in the text.

We shall estimate the probability that this branch is not certifying
that $w$ does not occur anywhere within the block.  As all
certificates are negative if $|I|<m$, this implies in particular that
$I$ is not a certificate, thus, in such a case, we know that the
algorithm shall continue, and this branch will contribute to a lower
bound to the complexity (both ordinary and certificate) of the given
algorithm.

The word $w$ has $d:=L-m+1$ potential occurrence positions within the
block. The occurrence at position $j+1$ may still occur if
$\xi_{i_a}=w_{i_a-j}$ for all $a\leq t$ such that
$1 \leq i_a-j \leq m$.  Thus, if we average na\"ively over all words
of length $m$, this gives a probability at least $s^{-t}$ (it may be
larger if some of the positions $i_a-j$ are out of range). Annoyingly,
the probabilities for different values of $j$ are
correlated. Nonetheless, a lemma proven by Yao in \cite{yao} (called
the \emph{Counting Lemma}) states that, among the $d$ possible
positions, there exists a set $J$, of cardinality at least
$\lceil d/t^2 \rceil$, for which the values 
$\{i_a-j\}_{a\leq t,\,j \in J}$ are all distinct, and thus, when
averaging over one word, the probabilities decorrelate. For the
positions $j \not\in J$, we are neglecting the possibility that they
are enforcing $I$ not to be a certificate, and thus we are
under-estimating the depth of certificate nodes, which gives a valid
lower bound.

In the case of multiple words the things are not harder.  When
averaging over all dictionaries of a given content, with repetitions
allowed, the probabilities associated to different words just
decorrelate, so that we do not need to generalise the counting lemma
of Yao in passing from his analysis of single-word dictionaries to our
context of arbitrary dictionaries.

If, as we do here, we insist on considering only negative
certificates, we can drop from the dictionaries all the words of
length smaller than $t$, i.e.\ consider the content $\vr'$ defined as
$r'_m=r_m$ if $m>t$ and $r'_m=0$ otherwise (of course, this makes an
even smaller set $X$ w.r.t.\ (\ref{eq.defXL})).  The overall number of
pairs of word/position which are certified to be independent for a
given set $I$ of length $t$ is thus at least a certain quantity $c_0\,
s^t$, regardless from the precise set $I$, where $c_0$ is defined as
\be
c_0 = c_0(\vr',L,t) = s^{-t} 
\sum_{m>t} r_m \left\lceil \frac{L+1-m}{t^2} \right\rceil
\ef,
\label{eq.defcstart}
\ee
where we make a crucial use of Yao's Counting Lemma.  Given that we
aim at producing a bound in terms of the function $\phi(\vr)$, only
the value $m^* := \argmax_m \frac{\ln s\,m\,r_m}{m}$ is relevant. So
we expect that the following analysis will show that the optimal value
of $t$ in (\ref{eq.lb14452bis}) is smaller than $m^*$, and that at
leading order we would obtain the same estimate of the complexity if,
instead of $c_0$, we use the smaller value
\be
c = s^{-t} r_{m^*} \left\lceil \frac{L+1-m^*}{t^2} \right\rceil
\ef.
\label{eq.defcp}
\ee
We are now ready to bound the quantity $f_{\vr}(I)$ in
(\ref{eq.defFvr}), of crucial importance in (\ref{eq.lb14452bis}), by
a function of $t$ and $c$ (or $c_0$). For any dictionary $D$, and at
the given value of $t$, we have a list $\cX$, of length $X=c\, s^t$,
of $t$-uples in $\Sigma^t$ corresponding to the entries of any given
word $w$, at its positions \hbox{$\{i_a-j\}_{a=1,\ldots,t}$,} for a pair
$(w,j)$ in the outcome of Yao's Counting Lemma. We also have the list
$\cY=\Sigma^t$, thus of length $Y=s^t$, of all possible $t$-uples,
i.e., of all possible texts restricted to the positions of $I$.  For a
pair $(w,j)$ in $\cX$, if $i_t-|w| \leq j \leq i_1-1$, call
$\psi(w,j) = (w_{i_1-j}, \ldots,w_{i_t-j})$. If instead we have some
$(i_a-j)$ out of range, define the corresponding entry of $\psi(w,j)$
by taking one character at random in the alphabet.  The map $\psi$ is
thus from $\cX$ to $\cY$, and, because of the forementioned
decorrelation, $\psi$ is distributed uniformly over all possible $Y^X$
maps, and $f_{\vr}(I)$ is the probability of the event that $\psi$ is
not a surjection.

It is well-known (see e.g.\ \cite[pag.~107]{flaj}) that the fraction
of random maps from a set of cardinality $X$ to a set of cardinality
$Y$ which are surjections is
\be
\label{eq.pStirl}
\pi_{X,Y}
=Y^{-X} Y! \left\{ \atopp{X}{Y} \right\}
\ee
where $\left\{ \atopp{n}{r} \right\}$ are the Stirling numbers of
second kind.  This quantity is used in the relation
\be
f_{\vr}(I)
\leq 1-\pi_{c\, s^t,s^t}
\ef,
\ee
valid for all sets $I$ of size $t$. We have an inequality because we
have produced lower-bound estimates on various quantities (for
example, the actual list associated to $I$ may be longer than $c\,
s^t$, or there could be entries \hbox{$(i_a-j)$'s} out of range, which
give rise to more pairs $(w,j)$ to be checked), and, importantly,
$\pi_{X,Y}$ is a monotone function of its first argument. As a result,
\be
\tilde{p}_t(\vr)
=
1-
\sum_I f_{\vr}(I)
\geq
1-\binom{L}{t} (1-\pi_{c\, s^t,s^t})
\ef.
\label{eq.ptil1}
\ee

%-------------------------------------------------------
\subsection{Estimates}

At this point we have completed our abstract reasonings, and we only
need to combine the inequalities
(\ref{eq.lb14452bis}),
(\ref{eq.defcp}),
% {eq.defcstart}),
(\ref{eq.pStirl}) and
(\ref{eq.ptil1}),
and to estimate them effectively. We shall start with the
forementioned fraction of surjections. We will use the fact that for
$c>1$ there exists a function
%$\tilde{c}(c)=c+\mathcal{O}(c e^{-c})$
$\tilde{c}(c)=c-\frac{1}{2}c \, e^{-c}+ \ldots$
such that 
$\pi_{cn,n} \geq \exp[-n e^{-\tilde{c}}]$ for all $n$. 
The precise function is
\be
\label{eq.4765738546}
\tilde{c}(c)
=
-\ln \big[
c - c \ln(c) - 
\ln( -1 + e^{c + W(-c \, e^{-c})} ) +
c \ln( c + W(-c \, e^{-c}) ) \big]
\ee
where $W(w)$, the Lambert $W$-function, is the principal solution for
$w$ in $z=w e^w$.  This is discussed, e.g., in 
\cite[Lemma 6 and Remark 7]{BasNic07}).
As a matter of fact, investigating (\ref{eq.4765738546}) gives the
two-sided bound
\be
\label{eq.ctilbou}
1-
% \frac{1}{c^2}
e^{-c+1} 
\leq \frac{\tilde{c}(c)}{c} \leq 1
\ef,
\qquad c \geq 1
\ee
or also the simpler but weaker
\be
\label{eq.ctilbou2}
1-\eps_1
% \frac{1}{c^2}
% e^{-c+1} 
\leq \frac{\tilde{c}(c)}{c} \leq 1
\ef,
\qquad c \geq 1 + \eps_2
\ee
for some pairs $(\eps_1,\eps_2)$, including, for example,
$(\eps_1,\eps_2)=(1/240,4)$.
%% An exact expression for $\pi_{A,B}$, easily deduced from the
%% definition and the inclusion-exclusion principle,
%% % (see e.g., \cite[]{AlonSpenc})
%% is as follows
%% \be
%% \pi_{A,B}
%% =
%% \sum_{r=0}^B
%% (-1)^r \binom{B}{r} \left(
%% \frac{B-r}{B} \right)^A
%% \ef.
%% \label{eq.5554365}
%% \ee
%% Their asymptotic estimate is well controlled, especially
%% in the simpler regime $c = A/B \gg 1$, which will correspond to our
%% case (see for example \cite[Lemma 6 and Remark 7]{BasNic07}).
%
We shall now analyse the resulting expression (\ref{eq.lb14452bis})
for the lower bound on $\Phi$. From this point on, 
%% we shall just use the expression for $c'$, defined in
%% (\ref{eq.defcp}), instead of $c$, although for brevity we shall use
%% $c$ and $\tilde{c}$ for $c'$ and $\tilde{c}(c')$,
for brevity we shall use $m$ for $m^*$, the argmax of the function
$\phi(\vr)$, and $r$ for $r_{m^*}$.  We shall also call for short
$Q=\ln (s\,m^*\,r_{m^*})=\ln (s\,m\,r)$, so that $\phi(\vr)=Q/m$.  We
thus have, neglecting round-offs,
\begin{align}
\label{eq.65465435}
c &= \frac{r (L+1-m)}{t^2 \,s^t}
\ef;
&
\tilde{p}_t &= 1- \binom{L}{t} (1-\pi_{c\, s^t,s^t})
% e^{-s^t e^{-\tilde{c}}})
\ef.
\end{align}
Monotonicity in $\tilde{p}$ gives that, under the condition
$\tilde{p}_t \leq 1 - (2 s\, \ka_s \ln s)^{-1}$, for some given
constant $\ka_s$, we could get the desired summand 
$1/(2 s\,\ka_s \,\mmin)$, on the LHS of (\ref{eq.3654635}), from the
second term alone of (\ref{eq.lb14452bis}).  So, in order to conclude
we could just replace (\ref{eq.lb14452bis}) by the simpler
% check that, for some value $t \geq L/\mmin$, 
\begin{align}
\frac{1}{\ln s}
\frac{L\, \phi(\vr)
}{\ka_s t}
% (t+1)}
\leq
\tilde{p}_t 
&\leq 
1 - \frac{1}{2 s\, \ka_s \ln s}
\ef.
\label{eq.u635464}
\end{align}
%% the term
%% %
%% $(\ln s)\, (t+1)\, \tilde{p}_t/L$ is at least as big as
%% $\phi(\vr)/(2\ka_s)$.  
%% %
In our parametrisation
% , and by choosing to work with the quantity $c'$, 
a condition holds for all possible dictionary contents $\vr$ if
it holds for $\mmin \geq 2$, $m \geq \mmin$ and $1 \leq r \leq s^m$
(recall that $r_m \leq s^m$),
% for all $m$ in the range of the content),
and this latter condition can be rephrased as $\ln(sm) \leq Q \leq
\ln(sm) + m \ln s$.
Define $\Omega_s \subseteq
% \mathbb{N} \times 
(\mathbb{R}^+)^2$ as
$\Omega_s=\{(m,Q) \;|\; m \geq 2 \,,\, 
\ln(sm) \leq Q \leq \ln(sm) + m \ln s \}$.
Then (\ref{eq.u635464}) can be rephrased as follows: for any $s\geq2$,
there exists a positive real value $\ka_s$ such that, for all $(m,Q)
\in \Omega_s$, there exist values $t$ and $L$ satisfying the
forementioned constraints, with $c$ and $\tilde{p}_t$ as
in~(\ref{eq.65465435}).

%% ..............

%% The LHS inequality of (\ref{eq.u635464})
%% % condition 
%% % $(\ln s)\, (t+1)\, \tilde{p}_t/L \geq \phi(\vr)/(2\ka_s)$ above 
%% can be rephrased in the following slightly stronger condition: for any
%% $s\geq2$, there exists a positive real value $\ka_s$ such that, for
%% all $(m,Q) \in \Omega_s$, there exist values $t$ and $L$ such that
%% %
%% \be
%% \label{eq.487653746}
%% \frac{(\ln s)\, t
%% % (t+1)
%% \, \tilde{p}_t}{L}
%% \frac{2m\,\ka_s}{Q}
%% \geq 1
%% \ef,
%% \ee
%% %
%% with $c$ and $\tilde{p}_t$ as in (\ref{eq.65465435}), and such that
%% $\tilde{p}_t \leq 1 - (4 s\, \ka_s \ln s)^{-1}$.  

%% ..........

This problem is, in principle, just a matter of elementary
calculus. The only difficulty is the fact that the involved
expressions are quite cumbersome.
%% As a result, we have reached a point at which we just have to study
%% a function in a finite-dimensional space: the LHS of
%% (\ref{eq.487653746}) is
The use of the union bound in deriving (\ref{eq.lb14452bis2}) is a
very crude approximation, which does not even make clear the fact that
$p_t$ is a positive quantity (as its estimate $\tilde{p}_t \leq p_t$)
may be negative). We shall start by investigating under which
conditions we have, say,
$\tilde{p}_t \geq 1 - \delta$, for some $0<\delta<1$. This condition
reads
\be
\label{eq.764658765}
\binom{L}{t} (1-\pi_{c s^t,s^t}) \leq \delta
%\frac{1}{2}
\ee
which, by Stirling approximation, can be strenghtened into
\be
1-\pi_{c s^t,s^t} \leq 
\delta
% \frac{1}{2} 
\left( \frac{Le}{t} \right)^{-t}
\ee
By using 
$\pi_{c s^t,s^t} \geq \exp(-s^t e^{-\tilde{c}})$, and
$1-e^{-x} \leq x$,
% for $x>0$, 
this can be strenghtened
into
\be
s^t e^{-\tilde{c}}
\leq 
\delta
% \frac{1}{2} 
\left( \frac{Le}{t} \right)^{-t}
\ef,
\ee
that is, we shall verify the condition
\be
\tilde{c}
\geq 
-\ln \delta + t \ln
\left( \frac{Les}{t} \right)
\ef,
\ee
or, by using the bound (\ref{eq.ctilbou2}), its strenghtening
% CHANGE THIS IF REQUIRED........
\begin{align}
\label{eq.67653545}
(1- \eps_1) c
% c\, (1-e^{-c+1})
& \geq 
-\ln \delta + t \ln
\left( \frac{Les}{t} \right)
\ef;
&
c \geq 1+\eps_2
\ef.
\end{align}
At this point, we shall find values of $L$, $t$ and $\delta$
satisfying both (\ref{eq.67653545}) and the two sides of
(\ref{eq.u635464}) (but in the simplifying situation in which we use
$\tilde{p}_t \geq 1 - \delta$ instead of (\ref{eq.65465435})).

%% i.e.\ those maximising our estimate for 
%% %
%% $\Phi_{\rm cert}(\vr)$, which is related, to a certain extent, to
%% maximising the difference between the LHS and RHS of
%% (\ref{eq.67653545}).  

Instead of solving a maximisation problem, we provide an \emph{ansatz}
for values $(L,t,\delta)$ which are almost optimal for the given
$(s,m,r)$.  To start with, we shall choose to reparametrise the
dependence from $L$ and $t$ into a dependence from two more convenient
new variables, $a$ and $\tau$, defined by
\begin{align}
L&=\frac{Q+a}{Q} m
\ef;
&
t&= \tau \frac{Q+a}{\ln s}
\ef.
\end{align}
We have a range $a, \tau \geq 0$, accounting for $t \geq 0$ and 
$L \geq m$.
% , for some positive values $\tau$ and $a$
%, sublinear w.r.t.\ the scale of $Q$, 
% to be determined later on.
The left-most inequality in (\ref{eq.u635464}),
which implicitly gives a bound on $\phi(\vr)$ in terms of
$\tilde{p}_t$, $L$, $t$ and $s$,
% {eq.487653746}) 
considerably simplifies under this reparametrisation, and gives 
\be
\begin{split}
\label{eq.487653746bis}
\frac{(\ln s)\, t
%(t+1)
\, \tilde{p}_t}{L}
\frac{\ka m}{Q}
&\geq
\frac{(\ln s)\, t\, (1-\delta)}{L}
\frac{\ka m}{Q}
\\
&=
\frac{(\ln s)\, \frac{(Q+a) \tau}{\ln s}\, (1-\delta)}
{\frac{Q+a}{Q} m}
\frac{\ka m}{Q}
=
\tau \ka \,(1-\delta)
\geq 1
\ef,
\end{split}
\ee
while right-most inequality in (\ref{eq.u635464}), which is already
rather simple in form, reads
\be
\ka \geq \frac{1}{2 \delta s \ln s}
\ee
from which we deduce that the following choice for $\ka_s$ is legitimate
\be
\label{eq.bestkaPre}
\ka_s = \min_{(\tau,\delta) \in \overline{\Omega}_s}
\max \left( \frac{1}{(1-\delta) \tau}, \frac{1}{2 \delta s \ln s} \right)
\ee
where $\overline{\Omega}_s \subseteq [0,1]^2$ is the region of
parameters $(\tau,\delta)$ such that (\ref{eq.67653545}) is verified
(not to be confused with $\Omega_s$, which is the domain in $(m,Q)$ in
which (\ref{eq.67653545}) shall be verified).  The reparametrisation
is convenient also for the quantities entering (\ref{eq.67653545}),
and it gives
\begin{align}
c &=
\frac{e^{Q(1-\tau)}}{Q(Q+a)^2 \tau^2}
\frac{a e^{-\tau a} (\ln s)^2}{s}
\ef;
\label{eq.73454}
\\
t \ln
\left( \frac{Les}{t} \right)
&=
(Q+a) \tau 
\frac{
\ln \frac{me}{Q \tau \ln s}}{\ln s}
\ef.
\end{align}
The structure of equations (\ref{eq.67653545}) and (\ref{eq.73454})
shows the emergence of a natural barrier to obtain an arbitrarily
large lower-bound, as it should be the case. We can heuristically
observe that, in the limit of large $Q$, the exponential factor
$e^{Q(1-\tau)}$ in $c$ dominates over the algebraic powers of $Q$,
thus satisfying the inequality, \emph{provided that $\tau < 1$}. This
in turns gives a maximal value of $t$, which, from
(\ref{eq.lb14452bis}), gives a maximal value for the lower bound that
can be attained with the present methods.

Note that we have
\be
\ln \frac{me}{Q \tau \ln s}
=
\ln m - \ln \tau - \ln Q + \ln \frac{e}{\ln s}
\leq
Q - \ln \tau - \ln Q + \lam_s
\ef,
\label{eq.76454654}
\ee
with
$\lam_s = 1-\ln s-\ln \ln s$,
because $r \geq 1$.
%  (so that $Q> \ln m$). 
As a result we have
\be
\label{eq.8653654}
t \ln
\left( \frac{Les}{t} \right)
=
(Q+a) \tau 
\frac{
\ln \frac{me}{Q \tau \ln s}}{\ln s}
\leq
(Q+a) \tau 
\frac{Q - \ln \tau - \ln Q + \lam_s}{\ln s}
\ef.
\ee
Substituting (\ref{eq.73454}) and (\ref{eq.8653654}) into
(\ref{eq.67653545}) gives the slightly stronger
% \begin{align}
\be
\label{eq.67653545b}
\begin{split}
&
\frac{e^{Q(1-\tau)}}{Q(Q+a)^2 \tau^2}
\frac{a e^{-\tau a} (\ln s)^2}{s}
% c\, (1-e^{-c+1})
\geq 
\max \bigg[
1+\eps_2,
\\
&
\qquad\qquad
\frac{1}{1- \eps_1}
\left(
-\ln \delta + 
(Q+a) \tau 
\frac{Q - \ln \tau - \ln Q + \lam_s}{\ln s}
\right)
\bigg]
\ef;
\end{split}
\ee
%% \\
%% \frac{e^{Q(1-\tau)}}{Q(Q+a)^2 \tau^2}
%% \frac{a e^{-\tau a} (\ln s)^2}{s}
%% &
%% \geq
%% 1+\eps_2
%% \ef.
% \end{align}
% (\ref{eq.67653545b})\footnote{With $a=a(\tau,s)$ understood.} and
% (\ref{eq.67653545c})
% (\ref{eq.grossaQa})
This shall be verified for all $(m,Q) \in \Omega_s$, with values
$\eps_1$ and $\eps_2$ fixed, e.g., to $\eps_1=1/240$ and $\eps_2=4$,
which are a legitimate pair. However, the expressions depend on $Q$
but \emph{not} on $m$ (this has come as a result of the smarter
parametrisation in $a$ and $\tau$, and of some manipulations of the
bounds). As a consequence, instead of optimising for $(m,Q) \in
\Omega_s$, we shall just optimise for $Q \in [\ln (2s), +\infty [$.

At this point it is already clear that we will succeed in finding a
finite suitable value of $\kappa_s$ satisfying
(\ref{eq.bestkaPre}). Indeed, consider the case in which
$\delta(1-\delta) = \Theta(1)$ and $\tau = \Theta(\eps)$. We have
$\kappa_s=\Theta(\eps^{-1})< \infty$, and, in (\ref{eq.67653545b}),
\be
\label{eq.67653545bEps}
\begin{split}
&
\Theta(\eps^{-2})
\frac{e^{Q}}{Q(Q+a)^2}
\frac{a(\ln s)^2}{s}
\geq 
\max \bigg[
\Theta(1),
\\
&
\qquad\qquad
\left(
\Theta(1)
+ 
(Q+a)
\left(
\Theta(\eps)
\frac{Q 
- \ln Q + \lam_s}{\ln s}
+
\Theta(\eps \ln \eps)
\frac{1}{\ln s}
\right)
\right)
\bigg]
\\
&
%\qquad\qquad\qquad\qquad
\rule{220pt}{0pt}
Q \in [\ln (2s), +\infty [
\end{split}
\ee
which, at sight, is satisfied for $\eps$ small enough, even with an
arbitrary choice of $a$ finite. So, now it is just a matter of
providing explicit values, and trying to optimise them to some extent.

% ------------------------------

\begin{figure}[tb]
\begin{center}
\setlength{\unitlength}{0.45\textwidth}
\begin{picture}(1,0.8)
\put(0,0){\includegraphics[width=0.45\textwidth]{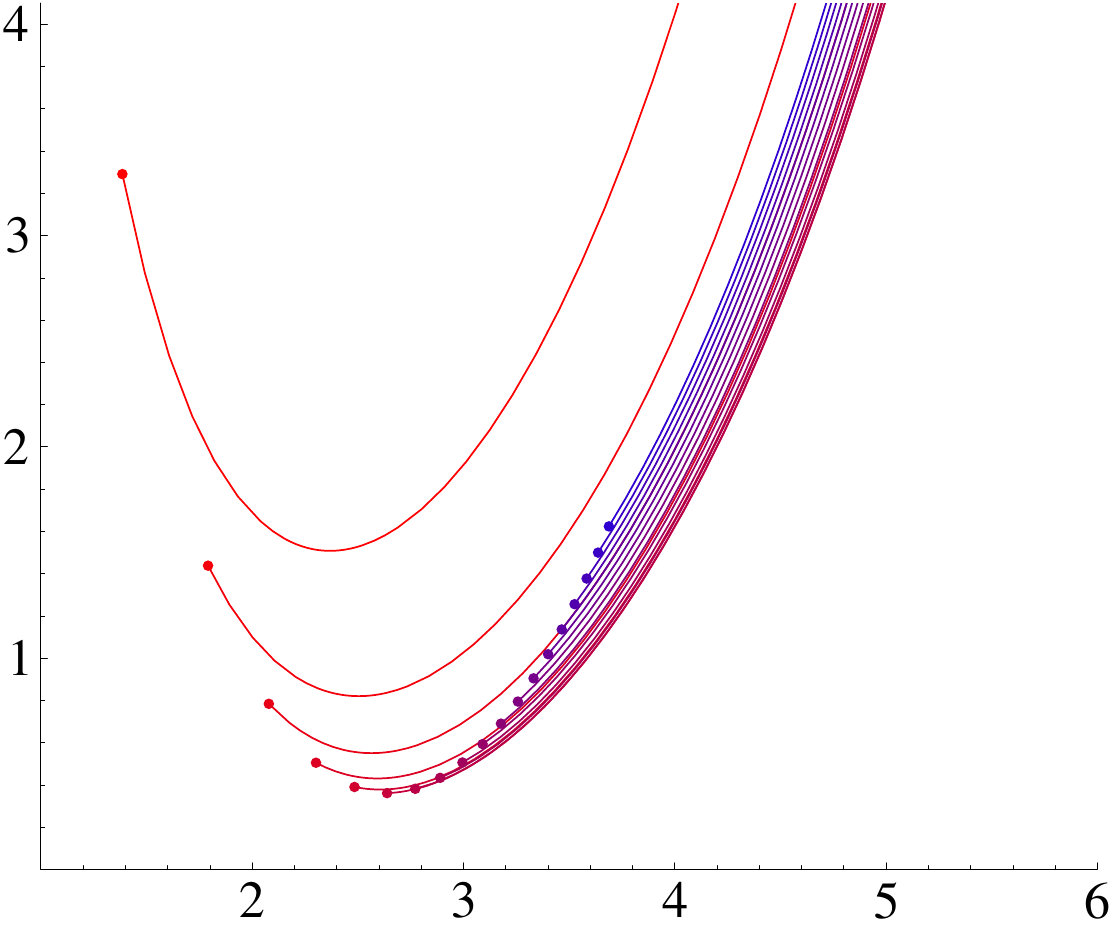}}
\put(.97,.08){$Q$}
\put(.06,.85){$f_s(Q)$}
\end{picture}
\qquad
\begin{picture}(1,0.8)
\put(0,0){\includegraphics[width=0.45\textwidth]{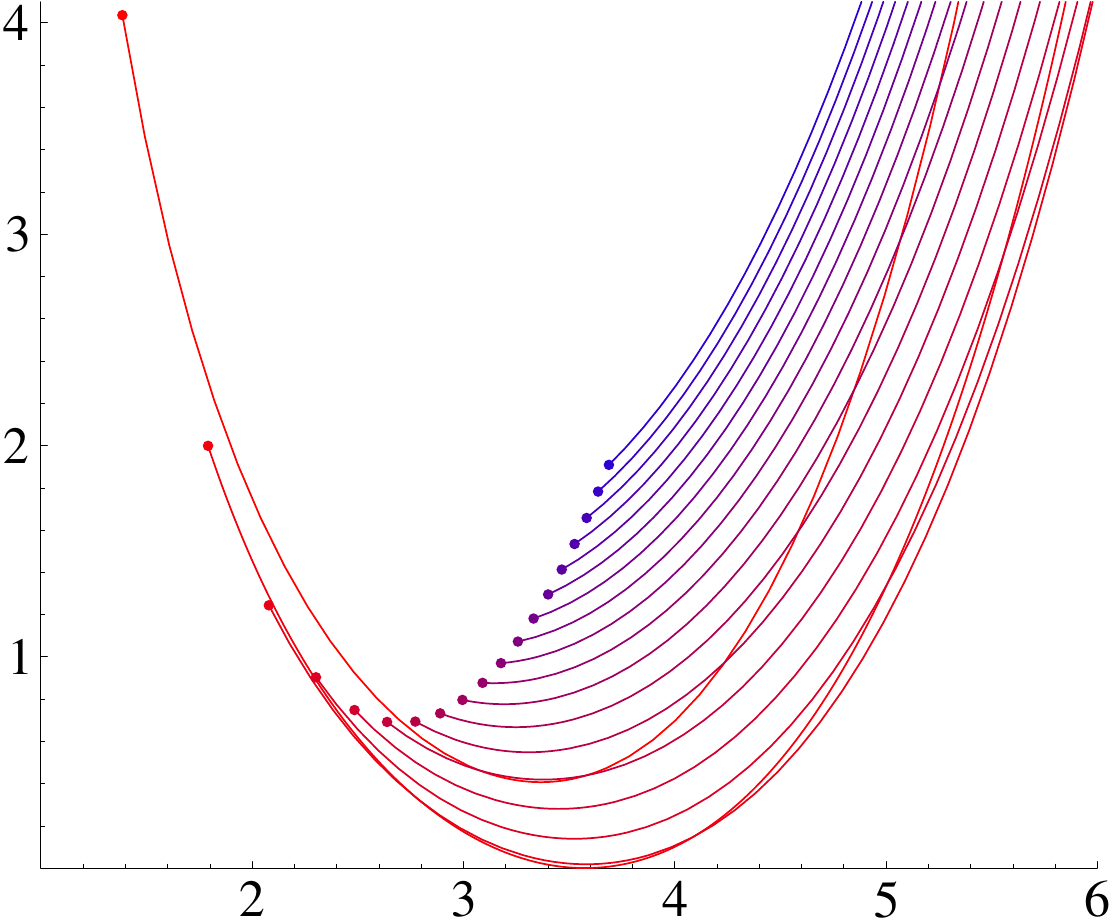}}
\put(.97,.08){$Q$}
\put(.06,.85){$g_s(Q)$}
\end{picture}
\end{center}
\caption{Curves $f_s(Q)$ and $g_s(Q)$, for $s=2,\ldots,20$ (red to
  blue), at the values $(T,\delta)=(0.21,0.03)$. The fact that all
  these curves remain above the real axis implies that
  $(T \frac{\ln s}{\sqrt{s}},\delta) \in \overline{\Omega}_s$ for all $s \geq 2$.
%  $2 \leq s \leq 20$ and $Q \in [\ln(2s),6]$.
\label{fig.curve}}
\end{figure}

Let us fix tentatively $a=Q/2$, and reparametrise $\tau$ as $\tau=T
\ln s/\sqrt{s}$.
% $\tau=T \,\ln s\, s^{-\frac{1}{2}}$.
This gives in (\ref{eq.67653545b}) (substitute $\eps_1=1/240$,
$\eps_2=4$, and use
$Q \geq \ln(2s)$)
\be
\label{eq.67653545c}
\begin{split}
&
\frac{2 e^{Q(1-T \frac{3 \ln s}{2 \sqrt{s}})}}{(3TQ)^2}
\geq 
\max \bigg[
5,
\frac{240}{239}
\left(
-\ln \delta + 
\frac{3TQ (Q - \ln T)}{2\sqrt{s}}
\right)
\bigg]
\ef.
\end{split}
\ee
It turns out that the inequality above is always satisfied for 
$s \in \{2,3,4,\ldots\}$, $Q \geq \ln(2s)$, $T \leq 0.212011$ and 
$\delta \geq 0.03613$, the worst case being $s=3$ and
$Q=3.57575\ldots$,
as indeed, for fixed values of $T$ and $\delta$, we
have a family (in $s$) of inequalities of the form
$a_s(Q)\geq \max \big( b_s(Q), c_s(Q) \big)$, which is satified if all
the functions $f_s(Q):=a_s(Q)-b_s(Q)$ and
$g_s(Q):=a_s(Q)-c_s(Q)$ are positive-valued in their ranges.
%
% $Q \in [\ln(2s),+\infty[$.
%
The behaviour for $Q$ or $s$ large is easily bounded, while
% at our tentative almost-optimal values of the parameters,
% $(T,\delta)=(0.21,0.03)$, 
the one near the worst-case region
$s,Q=\mathcal{O}(1)$, is illustrated in Figure~\ref{fig.curve}.  This
determines that $(T \frac{\ln s}{\sqrt{s}},\delta) \in
\overline{\Omega}_s$ for all $s \geq 2$. Thus, we can substitute the
corresponding values of $(\tau,\delta)$ in (\ref{eq.bestkaPre}), where
it turns out that
% and the values of $(T,\delta)$ provided above are such that
the max in (\ref{eq.bestkaPre}) is always attained on the left-most
quantity.

Using a round-off upper bound for the resulting constant
$\frac{1}{(1-\delta)T} = 
4.89354
% \frac{10000}{2037}
% 4.90918\ldots 
< 5$ 
finally gives the valid choice
\be
\label{eq.bestka}
\ka_s = 5 \frac{\sqrt{s}}{\ln s}
\ef.
\ee
This proves Proposition~\ref{prop.LB}.

The combination of Propositions \ref{prop.UB} and \ref{prop.LB}
clearly provides Theorem~\ref{thm1}.

%% ----------------------------------

%% OLDER CHOICE $a=Q$

%% Let us fix $a=Q$, and reparametrise $\tau$ as
%% $\tau=T \ln s/\sqrt{s}$.
%% % $\tau=T \,\ln s\, s^{-\frac{1}{2}}$.
%% This gives in (\ref{eq.67653545b}) (we still keep $\tau$ as a
%% shortcut, but substitute $\eps_1=7/6$, $\eps_2=1$, and use 
%% %
%% $Q \geq \ln(2s)$ at some points)
%% %
%% \be
%% \label{eq.67653545c}
%% \begin{split}
%% &
%% \frac{e^{Q(1-2T \frac{\ln s}{\sqrt{s}})}}{(2QT)^2}
%% \geq 
%% \max \bigg[
%% 2,
%% \frac{7}{6}
%% \left(
%% -\ln \delta + 
%% \frac{2TQ (Q - \ln T)}{\sqrt{s}}
%% \right)
%% \bigg]
%% \ef.
%% \end{split}
%% \ee
%% As a matter of fact, the inequality above is always satisfied for 
%% %
%% $s \in \{2,3,4,\ldots\}$, $Q \geq \ln(2s)$, $T \leq 1/5$ and 
%% %
%% $\delta >0.2888\ldots$, the worst case being $s=3$ and
%% $Q=4.65132\ldots$\ Substituting in (\ref{eq.bestkaPre}) and using a
%% rational upper bound of the resulting constant $3.51525\ldots$\ finally
%% gives the valid choice
%% %
%% \be
%% % \label{eq.bestka}
%% \ka_s = \frac{11}{3} \frac{\sqrt{s}}{\ln s}
%% \ef.
%% \ee

%%%%%%%%%%%%%%%%%%%%%%%%%%%%%%%%%%%%%%%%%%%%%%%%%%%%%%%
%%%%%%%%%%%%%%%%%%%%%%%%%%%%%%%%%%%%%%%%%%%%%%%%%%%%%%%
%\vspace*{-.4cm}

\appendix

\section{A simple bound on harmonic sums}
\label{app.546756}

Define
\be
S(x) := \sum_{m \geq x} \frac{1}{m^2}
\ef.
\ee
The main aim of this section is to determine the following fact.
\begin{lemma}
The function
\be
S^+_k(x) := \frac{k}{x} 
\Big( \zeta(2) - \sum_{m=1}^{k-1} \frac{1}{m^2}
\Big)
\ee
is an upper bound to $S(x)$ in the domain $x \geq k$:
\be
S(x)
\leq
S^{+}_k(x)
\qquad
\forall
\ 
x - k \in \mathbb{N}
\ef.
\label{eq.97646546}
\ee
\end{lemma}
Clearly, we have in particular $S^{+}_k(k) = S(k)$.  The lemma
follows as an immediate corollary of the following proposition
providing the induction step.
\begin{proposition}
\label{prop.165544354}
If
\be
\label{eq.34564}
\sum_{m \geq x} \frac{1}{m^2}
\leq \frac{A}{x}
\ee
then
\be
\sum_{m \geq x+1} \frac{1}{m^2}
\leq \frac{A}{x+1}
\ef.
\ee
\end{proposition}
This proposition, in turns, will be proven by using standard facts on
the monotone transport of measures, which we remind here.
\begin{definition}
\label{def.653654}
Let $\mu(k)$, $\mu'(k)$ be two normalised probability measures on
$\mathbb{N}$. We say that $\mu'$ is the \emph{monotone transport} of
$\mu$ if there exists a matrix $B$, lower-triangular and
left-stochastic, such that $\mu'(h) = \sum_{k} B_{hk}\, \mu(k)$.
\end{definition}
Then we have
\begin{proposition}
\label{prop.2654354}
Let $f(k)$ a real-valued function on $\mathbb{N}$. Define
\be
\left\langle f \right\rangle_{\mu} := \sum_k f(k) \mu(k)
\ef.
\ee
If $f$ is a monotone decreasing function, and $\mu'$ is the monotone
transport of $\mu$, then 
$\left\langle f \right\rangle_{\mu'} \leq \left\langle f
\right\rangle_{\mu}$.
\end{proposition}
{\it Proof.}
The matrix $B$ in Definition \ref{def.653654} has the property 
$B_{hk} \in \mathbb{R}^+$, $B_{hk}=0$ if $k>h$, and $\sum_h B_{hk}=1$
for all $k$. Thus we have, on one side,
\be
\left\langle f \right\rangle_{\mu}
=
\sum_k f(k) \mu(k) = 
\sum_{k\leq h} f(k) B_{hk} \,\mu(k)
\ee
and on the other side
\be
\left\langle f \right\rangle_{\mu'}
=
\sum_h f(h) \mu'(h) = 
\sum_{k\leq h} f(h) B_{hk} \,\mu(k)
\ee
so that
\be
\left\langle f \right\rangle_{\mu} 
- \left\langle f \right\rangle_{\mu'}
=
\sum_{k\leq h} \big( f(k)-f(h) \big) B_{hk} \,\mu(k)
\ee
As $B_{hk}$, $\mu(k)$ and $f(k)-f(h)$ are separately non-negative, the
latter because of the monotonicity of $f$, all the terms in the sum
are non-negative, and the statement follows.
\qed

We can now pass to the proof of the main statement.

\noindent
{\it Proof of Proposition \ref{prop.165544354}.}
Rewrite (\ref{eq.34564}) as
\be
A \geq 
\frac{\sum_{m \geq x} \frac{1}{m(m+1)} \frac{m+1}{m} }
{\sum_{m \geq x} \frac{1}{m(m+1)}}
\ee
If we define $\mu_x(m)$ as the probability measure on the integers
$\mu_x(m)=\frac{x}{m(m+1)} 1_{m \geq x}$, we thus have
\be
A \geq \eee_{\mu_x} \Big(\frac{m+1}{m} \Big)
\ef.
\ee
The measure $\mu_{x+1}$ is such that
$\mu_{x+1}(m) > \mu_{x}(m)$ if $m>x$ and
$\mu_{x+1}(m) = 0 < \mu_{x}(m)$ if $m=x$.
As a result, $\mu_{x+1}$ is the monotone transportation of
$\mu(x)$. The matrix $B$ is easily calculated\footnote{$B_{mm}=1$ for
  $m>x$, $B_{mx}=\frac{x+1}{m(m+1)}$ for $m>x$, all other entries are
  zero.}, although irrelevant at our purposes.
This fact, together with the monotonicity of the function
$\frac{m+1}{m}$, by Proposition \ref{prop.2654354} imply 
\be
\eee_{\mu_x} \Big(\frac{m+1}{m} \Big)
\geq
\eee_{\mu_{x+1}} \Big(\frac{m+1}{m} \Big)
\ee
and thus Proposition \ref{prop.165544354}.
\hfill $\square$

\vspace{2mm}
\noindent
The treatment above can be generalised. For $s>0$, define
\be
S_s(x) := \sum_{m \geq x} \frac{1}{m^{1+s}}
\ef.
\ee
We now prove the more general lemma
\begin{lemma}
The function
\be
S^+_{s,k}(x) := 
\left(
\zeta(1+s)-\sum_{m=1}^{k-1} \frac{1}{m^{1+s}}
\right)
\frac{\Gamma(k+s)\Gamma(x)}{\Gamma(k)\Gamma(x+s)}
\ee
is an upper bound to $S_s(x)$ for all
$x - k \in \mathbb{N}$.
\end{lemma}
Again, this is clearly the case for $x=k$, where we have
$S^+_{s,k}(x)=S_{s}(x)$.  Then, an analogue of
Proposition 
\ref{prop.165544354}
provides the induction step.
\begin{proposition}
If
\be
\label{eq.34564bis}
\sum_{m \geq x} \frac{1}{m^{1+s}}
\leq \frac{A\, \Gamma(x)}{s\,\Gamma(x+s)}
\ee
then
\be
\sum_{m \geq x+1} \frac{1}{m^{1+s}}
\leq \frac{A\, \Gamma(x+1)}{s\,\Gamma(x+1+s)}
\ef.
\ee
\end{proposition}

\noindent
{\it Proof.}
Recall the classical identity
\be
\sum_{m \geq x} \frac{\Gamma(m)}{\Gamma(m+1+s)}
=\frac{\Gamma(x)}{s\, \Gamma(x+s)}
\ef.
\ee
Then, rewrite (\ref{eq.34564bis}) as
\be
A \geq \frac{
\frac{s\, \Gamma(x+s)}{\Gamma(x)}
\sum_{m \geq x} 
\frac{\Gamma(m)}{\Gamma(m+1+s)}
\frac{\Gamma(m+1+s)}{\Gamma(m) m^{1+s}}
}
{
\frac{s\, \Gamma(x+s)}{\Gamma(x)}
\sum_{m \geq x} 
\frac{\Gamma(m)}{\Gamma(m+1+s)}
}
\ef.
\ee
If we define $\mu_{s,x}(m)$ as the measure on the integers
\be
\mu_{s,x}(m)=
\frac{s\, \Gamma(x+s)}{\Gamma(x)}
\frac{\Gamma(m)}{\Gamma(m+1+s)}
1_{m \geq x}
\ef,
\ee
we thus have
\be
A \geq \eee_{\mu_{s,x}} \Big(
\frac{\Gamma(m+1+s)}{\Gamma(m) m^{1+s}}
\Big)
\ef.
\ee
As above, the measure $\mu_{s,x+1}$ is such that $\mu_{s,x+1}(m) >
\mu_{s,x}(m)$ if $m>x$ and $\mu_{s,x+1}(m) = 0 < \mu_{s,x}(m)$ if
$m=x$, so, again, $\mu_{s,x+1}(m)$ is the monotone transportation of
$\mu_{s,x}(m)$.  The second crucial ingredient is the fact that the
function $\frac{\Gamma(m+1+s)}{\Gamma(m) m^{1+s}}$ is indeed
monotonically decreasing in $m$, for any value $s>0$.
% , which is a well-known fact on the $\Gamma$ function. 
As a result, again by using
Proposition \ref{prop.2654354}, we get
\be
\eee_{\mu_{s,x}} \Big(
\frac{\Gamma(m+1+s)}{\Gamma(m) m^{1+s}}
\Big)
\geq
\eee_{\mu_{s,x+1}} \Big(
\frac{\Gamma(m+1+s)}{\Gamma(m) m^{1+s}}
\Big)
\ee
and thus the proposition.
\hfill $\square$

%% ------------------

%% The aim of this section is to find constants $a_{\pm}(k)$,
%% $b_{\pm}(k) \in \mathbb{Q}$ such that the functions
%% \be
%% S^{\pm}_k(x) := \frac{1}{x} 
%% \big( \pi^2 a_{\pm}(k) + b_{\pm}(k) \big)
%% \ee
%% are upper/lower bounds to $S(x)$ in the domain $x \geq k$,
%% \be
%% S^{-}_k(x)
%% \leq
%% S_k(x)
%% \leq
%% S^{+}_k(x)
%% \qquad
%% \forall
%% \ 
%% x \geq k
%% \ef.
%% \ee
%% % A natural choice is to set
%% Remark that
%% \be
%% S_k(x) = S_k(k) -
%% \sum_{m=k}^{x-1}
%% \frac{1}{m^2}
%% \ef,
%% \ee
%% %
%% the sum being a non-negative quantity whenever $x \geq k$. Using
%% $\frac{1}{m(m+1)}\leq \frac{1}{m^2} \leq \frac{1}{m(m-1)}$ and
%% $\frac{1}{m(m-1)} = \frac{1}{m-1} - \frac{1}{m}$ allows to write
%% \be
%% -\frac{1}{k-1}+\frac{1}{x-1}
%% \leq
%% S_k(x) - S_k(k)
%% \leq
%% -\frac{1}{k}+\frac{1}{x}
%% \ee
%% so we can choose
%% \begin{align}
%% S_k^+(x) &= S_k(k)
%% -\frac{1}{k}+\frac{1}{x}
%% \ef;
%% &
%% S_k^-(x) &= S_k(k)
%% -\frac{1}{k-1}+\frac{1}{x-1}
%% \ef.
%% \end{align}
%% In particular,
%% \be
%% S_k^+(x) = 
%% \frac{1}{k}
%% \big( \pi^2 \frac{k}{6} - k \sum_{m=1}^{k-1} \frac{1}{m^2} \big)
%% -\frac{1}{k}+\frac{1}{x}
%% \ee

%% ---

%% \be
%% S^{\pm}_k(k) \equiv S_k(k)
%% =
%% \frac{1}{k}
%% \big( \pi^2 \frac{k}{6} - k \sum_{m=1}^{k-1} \frac{1}{m^2} \big)
%% \ee

%-------------------------------------------------------
% \section{Asymptotics of Stirling numbers of second kind}

% \vspace*{-.35cm}

%%%%%%%%%%%%%%%%%%%%%%%%%%%%%%%%%%%%%%%%%%%%%%%%%%%%%%%

\end{document}